\documentclass[aps,superscriptaddress,tightenlines,nofootinbib]{revtex4}
\usepackage{graphicx,amssymb,amsbsy,bm,amsmath}
\usepackage{hyperref}
\newcommand{\vx}{\ensuremath{\vec{x}}}
\newcommand{\vk}{\ensuremath{\vec{k}}}
\newcommand{\vq}{\ensuremath{\vec{q}}}

\newcommand{\bnu}{\ensuremath{\bar{\nu}}}

\newcommand{\be}{\begin{equation}}
\newcommand{\ee}{\end{equation}}
\newcommand{\bea}{\begin{eqnarray}}
\newcommand{\eea}{\end{eqnarray}}
\begin{document}
\title{Quantum corrections to slow roll inflation and
new scaling of superhorizon fluctuations.}
\author{D. Boyanovsky}
\email{boyan+@pitt.edu} \affiliation{Department of Physics and
Astronomy, University of Pittsburgh, Pittsburgh, Pennsylvania 15260,
USA} \affiliation{Observatoire de Paris, LERMA. Laboratoire
Associ\'e au CNRS UMR 8112.
 \\61, Avenue de l'Observatoire, 75014 Paris, France.}
\affiliation{LPTHE, Universit\'e Pierre et Marie Curie (Paris VI) et
Denis Diderot (Paris VII), Laboratoire Associ\'e au CNRS UMR 7589,
Tour 24, 5\`eme. \'etage, 4, Place Jussieu, 75252 Paris, Cedex 05,
France}
\author{H. J. de Vega}
\email{devega@lpthe.jussieu.fr} \affiliation{LPTHE, Universit\'e
Pierre et Marie Curie (Paris VI) et Denis Diderot (Paris VII),
Laboratoire Associ\'e au CNRS UMR 7589, Tour 24, 5\`eme. \'etage, 4,
Place Jussieu, 75252 Paris, Cedex 05,
France}\affiliation{Observatoire de Paris, LERMA. Laboratoire
Associ\'e au CNRS UMR 8112.
 \\61, Avenue de l'Observatoire, 75014 Paris, France.}
\affiliation{Department of Physics and Astronomy, University of
Pittsburgh, Pittsburgh, Pennsylvania 15260, USA}
\author{N. G. Sanchez}
\email{Norma.Sanchez@obspm.fr} \affiliation{Observatoire de Paris,
LERMA. Laboratoire Associ\'e au CNRS UMR 8112.
 \\61, Avenue de l'Observatoire, 75014 Paris, France.}
\date{\today}
\begin{abstract}
Precise cosmological data from WMAP and forthcoming cosmic microwave
background experiments motivate the study  of the quantum
corrections to the slow roll inflationary parameters. We find the
quantum (loop) corrections to the equations of motion of the
classical inflaton, to those for the fluctuations and to the Friedmann
equation in general single field slow roll inflation. We implement
a renormalized effective field theory (EFT) approach based on an
expansion in  $\left(H/M_{Pl}\right)^2$ and slow roll parameters
$\epsilon_V, \; \eta_V, \; \sigma_V, \; \xi_V$. We find that the
leading order quantum corrections to the inflaton effective
potential and its equation of motion are determined by the power
spectrum of scalar fluctuations. Its near scale invariance
introduces a strong infrared behavior naturally regularized by the
slow roll parameter $\Delta = \eta_V-\epsilon_V = \frac12(n_s -1)+
r/8 $. To leading order in the (EFT) and slow roll expansions we
find $ V_{eff}(\Phi_0)=
V_R(\Phi_0)\left[1+\frac{\Delta^2_{\mathcal{T}}}{32} \frac{n_s -1 +
\frac38 \; r}{n_s -1 + \frac14 \; r} +\textmd{higher orders}\right]
$ where $n_s$ and $r = \Delta^2_{\mathcal{T}}/
 \Delta^2_{\mathcal{R}} $ are the  CMB observables that depend
 implicitly on $\Phi_0$ and
$V_R(\Phi_0)$ is the \emph{renormalized} classical inflaton
potential. This effective potential during slow roll inflation is
strikingly {\bf different} from the usual Minkowski space-time
result. We also obtain the quantum corrections to the slow roll
parameters in leading order. Superhorizon scalar field fluctuations
grow for late times $ \eta \to 0^- $ as $ |\eta|^{-1+\Delta-d_-} $
where $ d_- $ is a {\bf novel quantum correction} to the scaling
exponent related to the \emph{self-decay} of superhorizon inflaton
fluctuations $\varphi \rightarrow  \varphi \varphi$ and $ \eta $ is
the conformal time. We find to leading order $
-d_-=\Delta_{\mathcal{R}}^2 \frac{\sigma_V \; (\eta_V-\epsilon_V) +
6 \, \xi^2_V}{4 \, (\eta_V-\epsilon_V)^2}$ in terms of the CMB
observables. These results are generalized to the case of the
inflaton interacting with a light scalar field $\sigma$ and we
obtain the decay rate $\Gamma_{\varphi \rightarrow \sigma \sigma}$.
These quantum corrections arising from interactions will compete
with \emph{higher order} slow-roll corrections in the gaussian
approximation and must be taken into account for the precision
determination of inflationary parameters extracted from CMB
observations.
\end{abstract}

\pacs{98.80.Cq,05.10.Cc,11.10.-z}

\maketitle

\section{Introduction}
Inflation is a  central part of early Universe cosmology passing
many observational tests and  becoming a predictive scenario
scrutinized by current and forthcoming observations. Originally,
inflation was introduced as an elegant explanation for several
shortcomings of the standard Big Bang
cosmology\cite{guth}-\cite{riottorev}. However, one of the most
compelling aspects of inflation is that it provides a mechanism for
generating scalar (density) and tensor (gravitational wave)
perturbations\cite{mukhanov}-\cite{bran}. A distinct aspect of
inflationary perturbations is that these are generated by quantum
fluctuations of the scalar field(s) that drive inflation. After
their wavelength becomes larger than the Hubble radius, these
fluctuations are amplified and grow, becoming classical and
decoupling from  causal microphysical processes. Upon re-entering
the horizon, during the matter era, these classical perturbations
seed the inhomogeneities which generate structure upon gravitational
collapse\cite{mukhanov}-\cite{bran}. While there is a great
diversity of   inflationary models, most of them predict fairly
generic features: a gaussian, nearly scale invariant spectrum of
(mostly) adiabatic scalar and tensor primordial fluctuations. These
generic predictions of most inflationary models make the
inflationary paradigm fairly robust. The gaussian, adiabatic and nearly
scale invariant spectrum of primordial fluctuations provide an
excellent fit to the highly precise wealth of data provided by the
Wilkinson Microwave Anisotropy Probe
(WMAP)\cite{komatsu,spergel,kogut,peiris}. Perhaps the most striking
validation of inflation as a mechanism for generating
\emph{superhorizon} (`acausal')  fluctuations is the anticorrelation
peak in the temperature-polarization (TE) angular power spectrum at
$l \sim 150$ corresponding to superhorizon
scales\cite{kogut,peiris}.

The confirmation of many of the robust predictions of inflation by
current high precision cosmological observations is placing
inflationary cosmology on solid grounds. Current and forthcoming
observations with ever increasing precision measurements will begin
to discriminate among different inflationary models, placing
stringent constraints on the underlying particle physics model of
inflation. There are small but important telltale discriminants
amongst different models: non-gaussianity, a running spectral index
for either scalar or tensor perturbations (or both), an isocurvature
component for scalar perturbations, different ratios for the
amplitudes between scalar and tensor modes, etc. Already WMAP
reports a hint of deviations from constant scaling exponents
(running spectral index)\cite{peiris}. Amongst the wide variety of
inflationary scenarios, \emph{slow roll}
inflation\cite{barrow,stewlyth} provides an appealing, simple and
fairly generic description of inflation. The basic premise of slow
roll inflation is that the  potential  is fairly flat during the
inflationary stage. This flatness not only leads to a slowly varying
Hubble parameter, hence ensuring a sufficient number of e-folds, but
also provides  an explanation for the gaussianity of the
fluctuations as well as for the (almost) scale invariance of their
power spectrum. A flat potential precludes large non-linearities in
the dynamics of the \emph{fluctuations} of the scalar field, which
is therefore determined by a gaussian free field theory. Furthermore,
because the potential is flat the scalar field is almost massless,
and modes cross the horizon with an amplitude proportional to the
Hubble parameter. This fact  combined with a slowly varying Hubble
parameter yields an almost scale invariant primordial power
spectrum.  Upon crossing the horizon the phases of the quantum
fluctuations freeze out and a growing mode dominates the dynamics,
i.e. the quantum fluctuations become classical (see ref.\cite{liddle} and
references therein). Departures from scale invariance and
gaussianity are determined by the departures from flatness of the
potential, namely by derivatives of the potential with respect to
the expectation value of the scalar field. These derivatives can be
combined into a hierarchy of dimensionless slow roll
parameters\cite{barrow} that allow an assessment of the
\emph{corrections} to the basic predictions of gaussianity and scale
invariance\cite{liddle}. This \emph{slow roll expansion}  has the
important bonus of allowing a reconstruction program that yields
details of the inflaton potential from observables extracted from
the analysis of CMB data, for example the index of the power spectra
of scalar and tensor perturbations, the ratio of their amplitudes,
etc.\cite{lidsey}. While more complicated scenarios can be proposed,
the current WMAP data seems to validate the simpler slow roll
scenario\cite{peiris}. Forthcoming precision CMB data forces a
deeper examination of the inflationary predictions, which has
motivated an analysis of the power spectra to higher order in the
slow roll expansion. A general slow roll
approximation\cite{stewart,domi} along with WKB\cite{martin,casa}
and uniform\cite{salman} approximations to study the power spectrum
beyond slow roll have been introduced. While progress is being made
in obtaining a more precise assessment of the power spectra of
scalar and tensor perturbations within the slow roll scenario, it
must be noted that all these refinements are still within the
\emph{gaussian} approximation, namely quadratic fluctuations of the
scalar field and the metric (or alternatively \emph{linear}
perturbations in the equations of motion for the fluctuations).
Interactions of the inflaton with other fields are a necessary
ingredient for a post-inflationary reheating stage where the energy
stored in the inflaton is transferred to other degrees of freedom
which eventually thermalize and lead to a transition from inflation
to the standard Hot Big Bang, radiation dominated cosmology.

Even within the simple single field slow roll scenario, higher
derivatives of the potential with respect to the homogeneous
expectation value of the scalar field will unavoidably lead to
non-linearities. The lowest order non-linearity results from a
\emph{cubic self-interaction} of the fluctuations around the
homogeneous expectation value. The strength of the cubic
self-interaction is determined by a particular slow-roll parameter,
(the `jerk' parameter)\cite{liddle,lidsey,peiris}.
This slow roll parameter also
enters in the running of the spectral index, and the current WMAP
data provides a rather loose bound on it\cite{peiris} which suggests
a small but non-vanishing cubic self-interaction.

Self-interactions of the fluctuations of the scalar field in turn
lead to \emph{non-gaussianities} which are characterized by a
non-vanishing \emph{bi-spectrum}\cite{allen}-\cite{7L}. The effect
of interactions on the decay of the inflaton in de Sitter space time
was  studied in refs.\cite{prem}, and  we have implemented a
dynamical renormalization group to study the decay of the quantum
fluctuations into other fields as well as the \emph{self decay} of
the fluctuations both for sub and super-horizon modes in slow roll
inflation\cite{ultimonuestro1,ultimonuestro2}.
In\cite{ultimonuestro2} the connection between the \emph{self-decay}
of inflaton
 fluctuations and the \emph{bi-spectrum} in single field slow roll
 inflation was established.

 In this article we study the effect of the  \emph{self interactions} 
as well as the
 interaction of the inflaton with other scalar fields to
assess the quantum corrections to the potential
  reconstruction program based on the slow-roll
 expansion. In particular, we study the quantum corrections to the equations
of motion of the expectation value of the inflaton, and to the fluctuations
as well as the quantum corrections to the Friedmann equation and to the slow
roll parameters. Such corrections are  important for an accurate assessment
of the inflationary parameters fit from the WMAP and future CMB data.

 \vspace{2mm}

\textbf{Inflation as an effective field theory:}  Effective field
theory provides a useful and physically motivated interpretation of
scalar field inflation below a cutoff scale. In this interpretation,
inflation is driven by a scalar field (the inflaton) with a fairly flat 
potential, which justifies the slow roll approximation and is consistent with
observational data. The inflaton replaces the microscopic description
provided by grand unified models in the cosmological space-time.
Such a description as an effective
field theory relies on a separation of scales, in this case  the
scale of inflation, determined by the Hubble parameter, and a high
energy scale $ M $. We identify $ M $ with  the Planck
scale $ M_{Pl} $ since so far, this is the only known energy scale above the
inflation scale. Within this effective field theory approach to
inflationary cosmology the inflaton model is   interpreted as   the
effective `low energy' field theory resulting from `integrating out'
the degrees of freedom with energy scales at or even above the scale $M$
(as advocated in refs.\cite{kalohol}).
In this interpretation, inflationary models are not fundamental
theories but {\it effective} descriptions in terms of a condensate,
the inflaton field. This type of description is very successful in a
wide variety of physically relevant cases: the low energy pion
dynamics emerging from full QCD and the Landau-Ginzburg effective
theories of superconductivity, superfluidity and critical phenomena.
In all these cases the low energy effective field theory allows a
systematic study of the \emph{universal} aspects of the relevant
dynamics. In this approach small dimensionless quantities are a
result of the ratio between the low and high energy scales. It is a
tantalizing possibility that the robustness of the predictions of
inflationary theories may be a manifestation of such `universality'
of the low energy effective field theory, akin to the robustness of
the description of  critical phenomena by  the Landau-Ginsburg
approach to phase transitions. Such point of view for inflationary
dynamics as an effective field theory driven by a scalar field has
been recently studied quantitatively in ref.\cite{hector}.

The small parameter that determines the validity of inflation as an
effective \emph{quantum field theory} below  the scale $M$ is $H/M $
where $H$ is the Hubble parameter during inflation and therefore the
scale at which inflation occurs. The slow roll expansion is in a
very well defined sense an \emph{adiabatic} approximation since the
time evolution of the inflaton field is slow on the expansion scale.
Thus the small dimensionless ratio $ H/M $, which is required for the
validity of an effective field theory (EFT) is logically
\emph{independent} from the small dimensionless combinations of
derivatives of the potential which ensure the validity of the
slow-roll expansion. In particular, since
 $ M=M_{Pl} $, the ratio $ H/M_{Pl} $ determines the amplitude
of tensor perturbations\cite{liddle}. Hence the validity of the
effective field theory description of inflation requires a very
small amplitude of tensor perturbations (gravitational waves) which
is consistent with the WMAP data\cite{peiris}.

Therefore, in this article we will invoke \emph{two independent}
approximations, the effective field theory (EFT) and the slow roll
approximation. The former is defined in terms of an expansion in the
ratio $ H/M_{Pl} $, whereas the latter corresponds to small slow roll
parameters.

In order to  determine the validity of the (EFT) and slow roll expansions,
it is important to highlight the main differences between slow roll
inflation and the post-inflationary stage. During slow roll
inflation the dynamics of the scalar field is slow on the time scale
of the expansion and consequently the change in the amplitude
of the inflaton is small and quantified by the slow roll parameters.
The slow roll approximation is indeed an \emph{adiabatic approximation}.
In striking contrast to this situation, during the post-inflationary stage of
reheating the scalar field undergoes rapid and large amplitude oscillations
that cannot be studied in a perturbative expansion\cite{reheatnuestro,ramsey}.
The slow roll approximation during inflation is warranted
because of the adiabatic evolution of the scalar field, and the (EFT)
is warranted because of the smallness of the ratio $ H/M_{Pl} $ as
determined by the WMAP data.

\vspace{2mm}

\textbf{The goals of this article:} In our previous work in
ref.\cite{ultimonuestro1,ultimonuestro2} we have found that the near
scale invariance of the inflaton fluctuations result in quantum
corrections that feature an \emph{infrared enhancement}.

The results of these references suggest that even when quantum
(loop) corrections are suppressed by powers of the effective field
theory ratio $ H/M_{Pl} $, there are enhancements arising
from infrared effects, a result of the near scale invariance of the
power spectrum of fluctuations.

In this article we focus our study on quantum aspects of
inflationary dynamics \emph{during the slow roll stage} by
considering quantum (loop) corrections from \emph{inflaton
fluctuations} as well as \emph{light} scalar fields with a nearly
scale invariant power spectrum. In particular we study the
contributions from the inflaton fluctuations and light scalar fields
to the \emph{effective inflaton potential} as well as self-energy
corrections to the equations of motion of the inflaton fluctuations
\emph{during slow roll}. We show that a {\bf strong infrared enhancement}
appears in the effective field theory when nearly scale invariant 
fluctuations are present. This is precisely the case in slow-roll
inflationary cosmology.

We restrict our study here to the quantum corrections from
fluctuations of the inflaton  and light scalar fields during slow
roll as a \emph{prelude} to a more complete study that should
eventually include gravitational fluctuations, a task that is
postponed for further study.

The motivation of this work is driven by forthcoming precision
measurements of the primordial power spectrum. These measurements
can potentially yield precise information to map out the
inflationary potential during inflation. In order to go from the
data to the inflaton potential and inflationary parameters, the
slow-roll approximation is typically invoked. Higher order
derivatives of the inflationary potential will be determined from
the slow roll parameters obtained from the data. The main point that
we  highlight in this article is that while the typical slow-roll
expansion is based solely on a free field description (gaussian) of
the fluctuations and a classical description of the inflaton
dynamics, quantum (loop) corrections yield contributions in the
slow-roll parameters.
These \emph{quantum} corrections compete with higher
order corrections in the slow roll approximation in the gaussian
theory and therefore  affect
the determination of the inflationary potential. Our study in this
article seeks to obtain a quantitative understanding of these
corrections. We find that the loop corrections have a
strong infrared behavior as a consequence of the nearly scale
invariant spectrum of scalar fluctuations. The infrared behavior is
manifest as poles in a small parameter $\Delta$ which is a measure
of the departure from scale invariance of the power spectrum of
scalar perturbations. $\Delta$ is a simple function of slow-roll
parameters.

\textbf{Brief summary of main results: }
\begin{itemize}
\item{We show that the one loop quantum correction to the equation
of motion for the homogeneous expectation value of the inflaton is
determined by the power spectrum of its quantum fluctuations. As
discussed in\cite{ultimonuestro2}, a nearly scale invariant power
spectrum of scalar fluctuations introduces a strong infrared
behavior: it is  naturally regulated by a small parameter $\Delta$
which measures the departure from scale invariance and is a simple
function of slow-roll parameters. The infrared divergences are
manifest as poles in $\Delta$\cite{ultimonuestro2}. We obtain the
effective equation of motion for the expectation value of the
inflaton field at one loop level in the effective field theory and
to leading order in the slow roll expansion. It is given by
$$
\ddot{\Phi}_0(t)+3\,H_0\,\dot{\Phi}_0(t)+
V^{'}_R(\Phi_0)\left\{1+ \frac{\Delta_{\mathcal{R}}^2}{n_s -1 +\frac{r}4}
\left[\frac{r}2 \left(n_s -1 +\frac{3 \, r}{16} \right) -
 \frac{dn_s}{d \ln k} \right]\right\} = 0 \; . $$}
\item{We obtain the one-loop corrections to the Friedmann equation
to leading order in the EFT and slow roll expansions. Just as for
the equation of motion for the expectation value of the inflaton,
the one-loop correction to the Friedmann equation is determined by
the power spectrum of the inflaton fluctuations and also features an
infrared enhancement. The effective inflaton potential is obtained
to leading order in the EFT and  slow roll approximations, it is
given by
$$
V_{eff}(\Phi_0) =  V_R(\Phi_0)\left[1+
\frac{r \; \Delta^2_{\mathcal{R}}}{32} \;
\frac{n_s -1 + \frac38 \; r}{n_s -1 + \frac14 \; r} \right] \; .
$$

where the CMB observables $n_s\,,\,r$ are implicit functions of
$\Phi_0$ through the slow roll parameters.  This effective potential
during slow roll inflation is strikingly different from the
effective potential in Minkowski space-time given by
eq.(\ref{potefM}). Moreover, the \emph{quantum corrections} to the
slow roll parameters are obtained in the leading order of the EFT
and slow roll expansions.}
\item{We obtain the renormalized equations of
motion for the superhorizon quantum fluctuations of the inflaton in
leading order in the EFT and slow roll expansions. The solution of
these equations features secular terms which are resummed via the
dynamical renormalization group\cite{ultimonuestro1,ultimonuestro2}.
This resummation reveals that superhorizon fluctuations display a
novel scaling dimension which is related to the \emph{self} decay of
inflaton fluctuations. We compute the quantum correction to the
scaling dimension and the rate for the \emph{self} decay of
superhorizon inflaton fluctuations in cosmic time, $\Gamma_{\varphi
\rightarrow \varphi\varphi}$. To leading order in (EFT) and slow
roll expansions we find for the correction to the scaling dimension

$$
-d_-=\Delta_{\mathcal{R}}^2 \frac{\sigma_V \; (\eta_V-\epsilon_V) +
6 \, \xi^2_V}{4 \, (\eta_V-\epsilon_V)^2}
$$
and for the \emph{self-decay} rate
$$
\Gamma_{\varphi \rightarrow  \varphi\varphi} = \frac12 \;
\Delta_{\mathcal{R}}^2 \; \frac{H_0 \; \xi^2_V}{(\eta_V-\epsilon_V)^2}
= \frac{2 \; H_0 \; \Delta_{\mathcal{R}}^2 \;
\xi^2_V}{\left(n_s -1 + \frac14 \; r\right)^{\! 2}} \; .
$$
These  results have been expressed in terms of CMB observables and
the jerk parameter $ \xi_V $ which is related  the running of the
index of scalar perturbations [see eq.(\ref{gorda})].}

\item{We generalize these results by studying a model in which the inflaton
interacts with
another light scalar field $\sigma$. We obtain the corrections to
the effective potential and scaling exponents from the self-energy
loop of $\sigma$ particles. In particular we obtain the partial
rate of superhorizon fluctuations of the inflaton decaying into two
$\sigma$ particles, $\Gamma_{\varphi\rightarrow \sigma \sigma}$.   }
\end{itemize}

One of the main points of this article is that quantum corrections
arising from the  interactions of the inflaton with itself and
\emph{any} other scalar field may compete with higher order
slow-roll corrections from the \emph{gaussian} approximation.

The article is organized as follows. In section \ref{eftsr} we
briefly discuss the systematics of the effective field theory (EFT)
and slow roll expansions. In section \ref{Eqofmotin} we obtain the
one loop correction to the equation of motion for the expectation
value of the inflaton. In this section we introduce the renormalized
effective field theory approach. Here we show that the
\emph{quantum} correction to the equation of motion is determined by
the power spectrum of the scalar fluctuations. The nearly scale
invariance of the power spectrum results in strong infrared behavior
of this contribution,  naturally regulated by the small parameter $\Delta$
which is a simple function of slow roll parameters. We obtain this
correction to leading order in the EFT and slow roll expansions. In
section \ref{friedeq} we obtain the one loop correction to the
Friedmann equation to leading order in the EFT and slow roll
expansions and identify the correct effective potential up to this
order. In this section we also obtain the \emph{quantum corrections}
to the slow roll parameters to leading order in the expansions. In
section \ref{anomdim} we obtain the equations of motion for the
quantum fluctuations of the inflaton including the one-loop self
energy. The solution of these equations feature secular terms that
are resummed by implementing the dynamical renormalization
group\cite{ultimonuestro1,ultimonuestro2}. The improved solution
features a quantum correction to the superhorizon scaling dimension,
which is obtained to leading order in the EFT and slow roll
expansions. In section \ref{other} we generalize the results to the
case in which the inflaton interacts with a light scalar  field. An
appendix shows that a calculation akin to that presented in sections
\ref{Eqofmotin}, \ref{friedeq} but in Minkowski space-time, lead to
the familiar effective potential. The Minkowski effective potential
is {\bf different} from the inflationary potential obtained in sec.
\ref{friedeq}. Section \ref{conclu} summarizes our results and
presents our conclusions.

\section{Effective field theory (EFT) and slow roll expansion.}\label{eftsr}

We consider single field inflationary models described by a
general self-interacting scalar field theory in a spatially flat
Friedmann-Robertson-Walker cosmological space time with scale
factor $a(t)$. In comoving coordinates the action  is given by
\begin{equation}\label{action}
S= \int d^3x \; dt \;  a^3(t) \Bigg[ \frac{1}{2} \;
{\dot{\phi}^2}-\frac{(\nabla \phi)^2}{2a^2}-V(\phi) \Bigg] \;.
\end{equation}
We consider a \emph{generic} potential $V(\phi)$, the only
requirement is that its \emph{derivatives} be small in order to
justify the slow roll expansion\cite{barrow,liddle,lidsey}.
In order to study the corrections from the quantum fluctuations we
separate the classical homogeneous expectation value of the scalar
field from the quantum fluctuations by writing
\be\label{tad}
\phi(\vec{x},t)= \Phi_0(t)+\varphi(\vec{x},t)\;,
\ee
\noindent with
\be\label{exp}
\Phi_0(t)=\langle \phi(\vx,t) \rangle~~;~~ \langle
\varphi(\vx,t)\rangle =0 \;,
\ee
where the expectation value is in the non-equilibrium quantum
state. Expanding the Lagrangian density and integrating by parts, the
action becomes
\be\label{Split}
S= \int d^3x \; dt \;  a^3(t)\,
\mathcal{L}[\Phi_0(t),\varphi(\vx,t)]\;,
\ee
\noindent with
\bea\label{lagra}
&&\mathcal{L}[\Phi_0(t),\varphi(\vx,t)]  =
\frac{1}{2} \; {\dot{\Phi}^2_0}-V(\Phi_0)+\frac{1}{2} \;
{\dot{\varphi}^2}-\frac{(\nabla \varphi)^2}{2 \, a^2} -\frac{1}{2}\;
V^{''}(\Phi_0)\; \varphi^2 \cr \cr &&- \varphi\;
\left[\ddot{\Phi}_0+3 \, H \,\dot{\Phi}_0+V^{'}(\Phi_0)\right] -
\frac{1}{6}\; V^{'''}(\Phi_0)\; \varphi^3 - \frac{1}{24}\;
V^{(IV)}(\Phi_0)\; \varphi^4+ \textmd{higher orders in}\,
\varphi \; .
\eea
We will obtain the equation of motion for the homogeneous
expectation value of the inflaton field  by implementing the tadpole
method (see \cite{ultimonuestro1,ultimonuestro2} and references
therein). This method consists in requiring the condition $\langle
\varphi(\vx,t)\rangle =0 $ consistently in a perturbative expansion
by treating the \emph{linear}, cubic, quartic (and higher order)
terms in the Lagrangian density eq.(\ref{lagra}) as
\emph{perturbations}\cite{ultimonuestro1,ultimonuestro2}.

Our approach relies on two distinct and fundamentally different
expansions: i) the effective field theory (EFT) expansion and  ii)
the slow-roll expansion.

\vspace{2mm}

{\bf The EFT expansion:} as mentioned above, the effective field
theory approach  relies on the separation between the energy scale
of inflation and the cutoff scale, which here is the Planck scale.
The scale of inflation is determined by the Hubble parameter during
the relevant stage of inflation when wavelengths of cosmological
relevance cross the horizon.  Therefore, the dimensionless ratio that
defines the EFT approximation is the ratio $H(\Phi_0)/M_{Pl}$,
where $H(\Phi_0)$ is the Hubble parameter during the relevant
inflationary stage. In scalar field driven inflation the reliability
of this approximation \emph{improves} upon dynamical evolution since
the scale of inflation {\it diminishes} with  time. Phenomenologically, the
EFT approximation is an excellent one since the amplitudes of
tensor and scalar perturbations $  \Delta_{\mathcal{T}} $ and
$ \Delta_{\mathcal{R}} $, respectively are given
by\cite{liddle,peiris}
\be\label{amps}
 \Delta_{\mathcal{T}} = \frac{\sqrt2}{\pi} \; \frac{H}{M_{Pl}}
\quad , \quad
\frac{H}{M_{Pl}} =  2 \, \pi \; \Delta_{\mathcal{R}} \;
\sqrt{2 \, \epsilon_V} \;,
\ee
\noindent where $\epsilon_V \ll 1$ is a slow-roll parameter (see
below). WMAP data\cite{peiris} yields $
\Delta_{\mathcal{R}} = 0.47 \times 10^{-4} $ thus providing strong
observational support to the validity of an effective field theory for
inflation well below the Planck scale and to the $ \frac{H}{M_{Pl}} $
expansion. Expected CMB constraints on $ \Delta_{\mathcal{T}} $
should still improve this observational support.

These perturbation amplitudes can be expressed in terms of the
semiclassical and quantum gravity temperature scales
\be\label{tem}
T_{sem} = \frac{\hbar \; H}{2 \, \pi \, k_B} \quad , \quad
T_{Pl} =  \frac{ M_{Pl} \; c^2 }{2 \, \pi \, k_B}
\ee
where $ k_B $ stands for the Boltzmann constant and $ c $ for the speed of
light. $ T_{sem} $ is the Hawking-Gibbons temperature of the initial
state (Bunch-Davis vacuum) of inflation while $ T_{Pl} $ is the Planck
temperature $\simeq 10^{32}$K. We have,
\be
 \Delta_{\mathcal{T}} =  \frac{\sqrt2}{\pi} \; \frac{T_{sem}}{T_{Pl}}
\quad , \quad \frac{T_{sem}}{T_{Pl}} = 2 \, \pi \; \Delta_{\mathcal{R}} \;
\sqrt{2 \, \epsilon_V} \;
\ee
Therefore, the WMAP data yield for the  Hawking-Gibbons temperature
of inflation: $ T_{sem} \simeq \sqrt{\epsilon_V} \; 10^{28}$K.

\vspace{2mm}

{\bf Slow roll approximation:} During slow roll inflation the
homogeneous expectation value $\Phi_0$ is a slowly varying
function of time which entails that the potential $V(\Phi_0)$ is
fairly flat as a function of $\Phi_0$. The slow roll expansion
introduces a hierarchy of small dimensionless quantities that are
determined by the derivatives of the potential.
Some\cite{barrow,liddle} of these (potential) slow roll parameters
are given by\footnote{We follow the definitions of $\xi_V;\sigma_V$
in ref.\cite{peiris}. ($\xi_V;\sigma_V$ are called  $\xi^2_V;\sigma^3_V$,
respectively, in\cite{barrow}).}
\bea
&&\epsilon_V  = \frac{M^2_{Pl}}{2} \;
\left[\frac{V^{'}(\Phi_0)}{V(\Phi_0)} \right]^2  \quad , \quad
\eta_V   = M^2_{Pl}  \; \frac{V^{''}(\Phi_0)}{V(\Phi_0)}\, ,
\label{etav} \\
&& \xi_V = M^4_{Pl} \; \frac{V'(\Phi_0) \;
V^{'''}(\Phi_0)}{V^2(\Phi_0)}  \quad , \quad  \sigma_V = M^6_{Pl}\;
\frac{\left[V^{'}(\Phi_0)\right]^2\,V^{(IV)}(\Phi_0)}{V^3(\Phi_0)}\;.
\label{sig}
\eea
The slow roll approximation\cite{barrow,liddle,lidsey} corresponds
to $\epsilon_V \sim \eta_V \ll 1$  with the hierarchy $\xi_V \sim
\mathcal{O}(\epsilon^2_V)~;~\sigma_V \sim
\mathcal{O}(\epsilon^3_V)$, namely $\epsilon_V$ and $\eta_V$ are
first order in slow roll, $\xi_V$ second order in slow roll, etc.

The slow roll parameters can be expressed in terms of the CMB observables
and their spectral runnings as follows,
\bea\label{gorda}
&&\epsilon_V  = \frac{r}{16} \quad , \quad \eta_V   =\frac12\left(
n_s - 1 +  \frac{3}{8} \, r \right) \quad , \quad 
\xi_V = \frac{r}4 \left(n_s - 1 +  \frac{3}{16} \, r \right)
- \frac12\frac{dn_s}{d \ln k}  \cr \cr
&&\sigma_V = - \frac{r}8 \left[ \left(n_s-1+\frac{r}{32}\right)^2 -
\frac{9 \, r^2}{1024} \right] + \frac14\left(n_s-1-\frac{9 \, r}8
+ \frac{r^2}{16} \right) \frac{dn_s}{d \ln k} \cr \cr
&&- \frac14\left(1 - \frac{r}6
 \right)  \left(n_s-1+\frac{3 \, r}8\right) \frac{dr}{d \ln k} +
\frac12 \left(1 - \frac{r}6  \right) \frac{d^2n_s}{d (\ln k)^2}
\eea
The Friedmann equation and the classical equation of motion for
$\Phi_0$ are
\bea
&&H_0^2 = \frac{1}{3 \, M^2_{Pl}}\left[\frac{1}{2}(\dot{\Phi}_0)^2+
V(\Phi_0)\right]\;, \label{hub} \\
\label{claseq} &&\ddot{\Phi_0}+3\,H_0\,\dot{\Phi}_0+V'(\Phi_0) =0
\;. \eea During slow roll inflation the equation of motion
(\ref{claseq}) can be approximated as \be\label{claseqsr}
\dot{\Phi}_0= -\frac{V'(\Phi_0)}{3\,H_0\,} + \textmd{higher orders
in slow roll}\,, \ee \noindent and  the Friedmann equation reads
\be\label{FRW} H_0^2 = \frac{V(\Phi_0)}{3 \,
M^2_{Pl}}\left[1+\frac{\epsilon_V}{3}+
\mathcal{O}(\epsilon^2_V,\epsilon_V \; \eta_V) \right]\;. \ee
 During slow roll, the number of e-folds before the end of
 inflation is given by
 \be\label{Nefold} N(\Phi) =
 \frac{1}{M_{Pl}}\int_{\Phi}^{\Phi_e}
\frac{d\Phi_0}{\sqrt{2 \, \epsilon_V(\Phi_0)}}\;,
 \ee
 \noindent where $\Phi_e$ is the value of $\Phi_0$ at the end of
 inflation. The stage of inflation during which wavelengths
 of cosmological relevance today first cross the Hubble radius
 corresponds to $N(\Phi) \sim  50$. Therefore, during this stage the
smallness of the slow
 roll parameter $\epsilon_V$ is justified by the large number of
 e-folds. In a wide variety of inflationary models  the slow roll
parameter $\epsilon_V$ is small for large $N(\Phi)$\cite{liddle,hector,peiris}.

We now introduce the effective mass of the fluctuations $M^2$  and
the cubic and quartic self-couplings $g,\lambda$ respectively as
\bea &&M^2 \equiv M^2(\Phi_0)  =   V''(\Phi_0) = 3 \; H_0^2 \;
\eta_V +
\mathcal{O}(\epsilon_V \; \eta_V)\,, \label{flucmass}\\
&& g\equiv g(\Phi_0)  =  \frac{1}{2} \;  V^{'''}(\Phi_0)\,,
 \label{g}\\
&& \lambda \equiv \lambda (\Phi_0)  =  \frac{1}{6} \;
 V^{(IV)}(\Phi_0)\,.\label{lambda}
\eea
In particular, the dimensionless combination of the cubic coupling
and the scale of inflation is given to leading order in slow-roll by
\be\label{gcoup}
\frac{g}{H_0} = \frac{3\,\xi_V}{2\sqrt{2\;\epsilon_V}} \; \frac{H_0}{M_{Pl}}
= \frac32 \, \pi \;  \Delta_{\mathcal{R}}\left[ \frac{r}2 \left(n_s - 1
+ \frac{3 \, r}{16} \right) - \frac{dn_s}{d \ln k}\right] \; ,
\ee
\noindent and the quartic coupling $\lambda$ can be conveniently
written in terms of slow-roll and effective-field theory parameters
as
\be\label{lam}
\lambda = \frac{\sigma_V}{4 \, \epsilon_V}\left(\frac{H_0}{M_{Pl}}\right)^2
= 2 \, \pi^2 \;  \Delta_{\mathcal{R}}^2 \; \sigma_V \; .
\ee
Moreover,  $\lambda$ can be written solely in terms of CMB observables
inserting the expression eq.(\ref{gorda}) for $ \sigma_V $ into eq.(\ref{lam}).

During slow roll the effective mass and couplings are not constants
but  \emph{very slowly varying functions of time}, and according
with the slow roll hierarchy, both the cubic and quartic self
couplings are small, the quartic being of higher order in slow roll
than the cubic etc, namely
$$
1\gg \frac{g}{H_0} \gg \lambda \gg \cdots
$$
\noindent where the dots stand for self-couplings arising from
higher derivative of the potential as displayed in eq.(\ref{lagra}).
The time dependence of these couplings is implicit through their
dependence on $\Phi_0$ determined by eq.(\ref{claseqsr}).
Eqs.(\ref{gcoup}) and (\ref{lam}) clearly show that $(g/H)^2$ and
$\lambda$ are of the \emph{same order} in the EFT expansion, namely
$\mathcal{O}(H^2_0/M^2_{Pl})$. This observation will be important in
the calculation of the self-energy correction for the quantum
fluctuations.

In order to keep a simple notation, the calculations will be
performed in terms of $g;\lambda$ [see definitions
eqs.(\ref{g})-(\ref{lambda})] and we will write these effective
couplings in terms of slow roll and EFT variables using
eqs.(\ref{gcoup})-(\ref{lam}) at the end of the calculations.

\section{Quantum corrections to the equation of motion for the
inflaton.}\label{Eqofmotin}

Quantum corrections to the equations of motion for the inflaton and
for the fluctuations will be
obtained by treating the second line in eq. (\ref{lagra}),
namely, the \emph{linear} and the non-linear terms in $\varphi$ in
perturbation theory.

The generating functional of non-equilibrium real time correlation
functions requires a path integral along a complex contour in time:
the forward branch corresponds to time evolution forward $(+)$ and
backward $(-)$ in time as befits the time evolution of a density
matrix. Fields along these branches are labeled $\varphi^+$ and
$\varphi^-$, respectively (see
refs.\cite{ultimonuestro1,ultimonuestro2} and references therein).
The tadpole conditions \be\label{tads} \langle \varphi^\pm(\vx,t)
\rangle =0 \; , \ee \noindent both lead to the (same) equation of
motion for the expectation value $\Phi_0(t)$ by considering the
\emph{linear, cubic} and higher order terms in the Lagrangian
density as interaction vertices. Up to one loop order we find
\be\label{1lupeqn} \ddot{\Phi}_0(t)+3 \, H \;
\dot{\Phi}_0(t)+V'(\Phi_0)+g(\Phi_0) \; \langle
[\varphi^+(\vx,t)]^2\rangle =0 \;. \ee The first three terms in
eq.(\ref{1lupeqn}) are the familiar ones for the equation of motion
of the inflaton.

The last term is the one-loop correction to the equations of motion
of purely quantum mechanical origin. Another derivation of this
quantum correction can be found in\cite{reheatnuestro,ramsey}.
The fact that the tadpole method, which in this case results in a one-loop
correction, leads to a covariantly conserved and fully renormalized energy
momentum tensor has been previously established in the most general case in
refs.\cite{reheatnuestro,erice} and more recently in ref.\cite{mottola}.

 The coupling $g$ is defined by
eq. (\ref{g}). The $\langle(\cdots)\rangle$ is computed in the free
field (Gaussian) theory of the fluctuations $\varphi$ with an
effective `mass term' $M^2$ given by eq. (\ref{flucmass}), the
quantum state will be specified below. Furthermore, it is
straightforward to see that $\langle [\varphi^+(\vx,t)]^2\rangle =
\langle [\varphi^-(\vx,t)]^2\rangle=\langle
[\varphi(\vx,t)]^2\rangle$. In terms of the spatial Fourier
transform of the fluctuation field $\varphi(\vx,t)$, the one-loop
contribution can be written as \be\label{lupPS} \langle
[\varphi(\vx,t)]^2\rangle = \int \frac{d^3 k}{(2\pi)^3} \; \langle
|\varphi_{\vk}(t)|^2 \rangle = \int_0^{\infty} \frac{dk}{k} \;
\mathcal{P}_{\varphi}(k,t)\,, \ee \noindent where $\varphi_{\vk}(t)$
is the spatial Fourier transform of the fluctuation field
$\varphi(\vx,t)$ and we have introduced the power spectrum of the
fluctuation \be\label{PS} \mathcal{P}_{\varphi}(k,t) = \frac{k^3}{2
\, \pi^2} \; \langle |\varphi_{\vk}(t)|^2 \rangle \,. \ee The metric
background is as usual,
$$
ds^2= dt^2-a^2(t) \; d{\vec x}^2 = C^2(\eta) \left[ (d \eta)^2 -  d{\vec x}^2
\right] \; ,
$$
where $ \eta $ is the  conformal time and $ C(\eta) \equiv a(t(\eta)) $.

In order to compute the one-loop contribution, it is convenient to
work in conformal time and to conformally rescale the field
\be\label{rescale}
\varphi(\vx,t) =\frac{\chi(\vx,\eta)}{C(\eta)}
\quad ,
\ee
\noindent $ C(\eta) $ being the scale factor in conformal time.

During slow roll inflation the scale factor is quasi de Sitter and
to lowest order in slow roll it is given by : \be\label{quasiDS}
C(\eta)=-\frac{1}{H_0 \; \eta} \; \frac{1}{1-\epsilon_V}=
-\frac{1}{H_0 \; \eta} (1+\epsilon_V) + \mathcal{O}(\epsilon_V^2)
\,. \ee The spatial Fourier transform of the free field Heisenberg
operators $\chi(\vx,\eta)$ obey the equation \be\label{heiseqn}
 \chi^{''}_{\vk}(\eta)+ \left[k^2 + M^2 \;  C^2(\eta)-
\frac{C^{''}(\eta)}{C(\eta)} \right]\chi_{\vk}(\eta)=0 \,.
\ee
Using the slow roll expressions eqs.(\ref{flucmass}) and (\ref{quasiDS}),
it becomes
\be\label{heiseqn2}
\chi^{''}_{\vk}(\eta)+ \left[k^2
-\frac{\nu^2-\frac{1}{4}}{\eta^2} \right]\chi_{\vk}(\eta)=0 \ee
\noindent where the index $\nu$ is given by \be\label{nu} \nu =
\frac{3}{2} + \epsilon_V-\eta_V
+\mathcal{O}(\epsilon^2_V,\eta^2_V,\epsilon_V\eta_V) \,.
\ee

The scale invariant case $ \nu = \frac{3}{2} $ corresponds to
massless inflaton fluctuations in the de Sitter background. The
quantity \be\label{delta} \Delta= \frac{3}{2}-\nu =
\eta_V-\epsilon_V +
\mathcal{O}(\epsilon^2_V,\eta^2_V,\epsilon_V\eta_V)\,. \ee measures
the departure from scale invariance. In terms of the spectral index
of the scalar adiabatic perturbations $ n_s $ and the ratio $ r $ of
tensor to  scalar perturbations, $ \Delta $ takes the form, \be
\Delta=\frac12 \left( n_s - 1 \right) + \frac{r}8 \; . \ee The free
Heisenberg field operators $\chi_{\vk}(\eta)$ are written in terms
of annihilation and creation operators that act on Fock states  as
\be\label{ope} \chi_{\vk}(\eta) = a_{\vk} \; S_{\nu}(k,\eta)+
a^{\dagger}_{-\vk} \; S^{*}_{\nu}(k,\eta) \ee \noindent where the
mode functions $S_{\nu}(k,\eta)$ are solutions of the eqs.
(\ref{heiseqn2}). We choose Bunch-Davis boundary conditions on these
mode functions so that \be\label{BDS} S_{\nu}(k,\eta) = \frac{1}{2}
\; \sqrt{-\pi\eta} \; e^{i\frac{\pi}{2}(\nu+\frac{1}{2})} \;
H^{(1)}_\nu(-k\eta)\, , \ee this defines the Bunch-Davis vacuum $
a_{\vk} |0>_{BD} =0 $.

There is no unique choice of an initial state, and a recent body of work
has began to address this issue\cite{inistate} (see ref.\cite{mottola}
for a discussion and further references). A full study of the
\emph{quantum loop} corrections with different initial
states must first elucidate the behavior of the propagators for the
fluctuations in such states. In this article we focus on the standard
choice in the literature\cite{liddle,lidsey} which allows us to include the
quantum corrections into the standard results in the literature. A study
of quantum loop corrections with different initial states is an important
aspect by itself which we postpone to later work.

The index $\nu$ in the mode functions eq.(\ref{BDS}) depends on the
expectation value of the scalar field,  via the slow roll variables,
hence it slowly varies in time. Therefore, it is  consistent to treat
this time  dependence of  $\nu$ as an \emph{adiabatic approximation}.
This is well known and standard in the slow roll
expansion\cite{liddle,lidsey}.
Indeed, there are corrections to the mode functions which are higher order
in slow roll as discussed in detail in refs.\cite{stewart}-\cite{salman}.
However, these mode functions enter in the propagators in loop corrections,
therefore they yield higher order contributions in slow roll
and we discard them consistently to lowest order in slow roll.

With this choice and to lowest order in slow roll, the power
spectrum eq.(\ref{PS}) is given by \be\label{PSSR}
\mathcal{P}_{\varphi}(k,t) = \frac{H^2}{8 \, \pi} \; (-k\eta)^3 \;
|H^{(1)}_\nu(- k \eta)|^2 \,. \ee For large momenta $|k\eta| \gg 1$
the mode functions behave just like free field modes in Minkowski
space-time, namely \be S_{\nu}(k,\eta) \buildrel{|k\eta| \gg
1}\over= \frac{1}{\sqrt{2k}} \; e^{-ik\eta} \ee \noindent Therefore,
the quantum correction to the equation of motion for the inflaton
eqs.(\ref{1lupeqn}) and (\ref{lupPS}) determined by the momentum
integral of $ \mathcal{P}_{\varphi}(k,t) $ features both quadratic
and logarithmic divergences. Since the field theory inflationary
dynamics is an \emph{effective field theory} valid below a comoving
cutoff $\Lambda$ of the order of the Planck scale, the one loop
correction (\ref{lupPS}) becomes \be\label{PSint} \int^{{\Lambda}}_0
\frac{dk}{k} \,\mathcal{P}_{\varphi}(k,t) = \frac{H^2}{8 \, \pi}
\int^{\Lambda_p}_0 \frac{dz}{z} \; z^3\,
\left|H^{(1)}_\nu(z)\right|^2 \,, \ee \noindent where $
\Lambda_p(\eta)$ is the ratio of the cutoff in physical coordinates
to the scale of inflation, namely \be\label{physcut} \Lambda_p(\eta)
\equiv\frac{\Lambda}{H \; C(\eta)}=-\Lambda\,  \eta\;. \ee The
integration variable $ z=-k \, \eta $ has a simple interpretation at
leading order in slow roll \be\label{zSR} z \equiv -k \, \eta =
\frac{k}{H_0 \, C(\eta)}= \frac{k_p(\eta)}{H_0} \,, \ee \noindent
where $k_p(\eta)=k/C(\eta)$ is the wavevector in physical
coordinates. If the spectrum of scalar fluctuations were strictly
scale invariant, (namely for massless inflaton fluctuations in de
Sitter space-time), then the index would be $\nu=3/2$ and  the
integrand in (\ref{PSint}) given by \be\label{integ} z^3 \,
\left|H^{(1)}_{\frac{3}{2}}(z)\right|^2 =
\frac{2}{\pi}\left[1+z^2\right]\,. \ee In this strictly scale
invariant case, the integral of the power spectrum also features an
\emph{infrared} logarithmic divergence. While the ultraviolet
divergences are absorbed by the renormalization counterterms in the
effective field theory, no such possibility is available for the
infrared divergence. Obviously, the origin of this infrared behavior
is the {\bf exact} scale invariance of superhorizon fluctuations.
However, during slow roll inflation there are small corrections to
scale invariance determined by the slow roll parameters, in
particular the index $\nu$ is slightly different from $3/2$ and this
slight departure from scale invariance introduces a natural infrared
regularization. In a previous article\cite{ultimonuestro2} we have
introduced an expansion in the parameter $\Delta = 3/2-\nu=
\eta_V-\epsilon_V+\mathcal{O}(\epsilon^2_V,\eta^2_V,\epsilon_V\eta_V)$
which is small during slow roll and we expect  here as in
ref.\cite{ultimonuestro2} that the infrared divergences featured by
the quantum correction manifest as \emph{simple poles} in $\Delta$.
We will now proceed to compute the quantum correction to the
equation of motion for the inflaton by isolating the pole in
$\Delta$ as well as the leading logarithmic divergences. To achieve
this goal we write the integral \be\label{intsplit}
\int^{\Lambda_p}_0 \frac{dz}{z} \; z^3 \; |H^{(1)}_\nu(z)|^2=
\int^{\mu_p}_0 \frac{dz}{z} \; z^3 \, \left|H^{(1)}_\nu(z)\right|^2
+ \int^{\Lambda_p}_{\mu_p} \frac{dz}{z} \; z^3 \,
\left|H^{(1)}_\nu(z)\right|^2 \, . \ee \noindent $\mu_p$ acts here
as  infrared cutoff for the first integral. The second integral is
ultraviolet and infrared finite for finite $\mu_p, \; \Lambda_p$. We
can set then $\nu=3/2$ in this integral and use eq. (\ref{integ}).
In the first integral we obtain the leading order contribution in
the slow roll expansion, namely the pole and leading logarithm, by
using the small argument limit of the Hankel functions. This yields,
\be z^3 \, \left|H^{(1)}_\nu(z)\right|^2 \buildrel{z \to
0}\over=\left[ \frac{2^{\nu} \; \Gamma(\nu)}{\pi} \right]^2 \; z^{2
\, \Delta} \ee and we find that eq.(\ref{intsplit}) yields after
calculation, \be\label{1int} \int^{\mu_p}_0 \frac{dz}{z} \; z^3 \,
\left|H^{(1)}_\nu(z)\right|^2 = \frac{2}{\pi}\left[\frac{1}{2 \,
\Delta}+ \frac{\mu^2_p}{2} + \gamma - 2 + \ln(2 \; \mu_p)
+\mathcal{O}(\Delta)\right]\,, \ee \noindent where we have displayed
the pole in $\Delta$ and the leading infrared logarithm. Combining
the above result with the second integral (for which we set
$\Delta=0$) we find the following final result for the quantum
correction to leading order in  slow roll
\be\label{QC}
\frac12 \langle[\varphi(\vx,t)]^2\rangle = \left(\frac{H_0}{4 \, \pi}\right)^2
\left[ {\Lambda_p}^2 + \ln \Lambda_p^2 +\frac1{\Delta}
 + 2 \, \gamma - 4 + \mathcal{O}(\Delta) \right]\,,
\ee
\noindent
where $\gamma$ is the Euler-Mascheroni constant.  While the
quadratic and logarithmic \emph{ultraviolet} divergences are
regularization scheme dependent, the pole in $\Delta$ arises from
the  infrared behavior and is independent of the regularization
scheme. In particular this pole coincides with that found
in the expression for $<\phi^2(\vx,t)>$ in ref.\cite{fordbunch}. The
\emph{ultraviolet divergences}, in whichever renormalization scheme,
 require that the effective field theory be
defined to contain \emph{renormalization counterterms} in the bare
effective lagrangian, so that these counterterms will systematically
cancel the divergences encountered in the calculation of quantum
corrections in the (EFT) and slow roll approximations.

\subsection{Renormalized effective field theory: renormalization counterterms}

The renormalized effective field theory is obtained by writing the
potential $V[\phi]$ in the Lagrangian density in (\ref{action}) in
the following form \be\label{count}
V(\phi)=V_R(\phi)+\delta\,V_R(\phi,\Lambda)\;, \ee \noindent where
$V_R(\phi)$ is the renormalized \emph{classical} inflaton potential
and $\delta\,V_R(\phi,\Lambda)$ includes the renormalization
counterterms which are found systematically in a slow roll expansion
by requesting that in the perturbative (slow roll) expansion
insertion of the counterterms cancel the ultraviolet divergences. In
this manner, the equations of motion and correlation functions in
this effective field theory \emph{will not depend on the cutoff
scale}. The counterterm required to cancel the ultraviolet
divergences in the inflaton equation of motion can be gleaned by
restoring the dependence of the coupling $g$ on $\Phi_0$, namely
\be\label{equalup}
\ddot{\Phi}_0(t)+3\,H\,\dot{\Phi}_0(t)+V'(\Phi_0)+ V^{'''}(\Phi_0)
\left(\frac{H_0}{4 \, \pi}\right)^2 \left[\Lambda_p^2+
\ln\Lambda_p^2 +\frac{1}{\Delta}+2 \, \gamma - 4 +
\mathcal{O}(\Delta)\right]=0 \,. \ee

>From this equation it becomes clear that the one-loop ultraviolet
divergences can be canceled by choosing \be\label{count2}
\delta\,V_R(\phi,\Lambda)=
\mathcal{C}_0[\Lambda,H_0]+\mathcal{C}_2[\Lambda,H_0] \;
V^{''}_R(\phi) +\textmd{higher orders in slow roll}\;, \ee \noindent
the extra terms refer to higher derivatives with respect to $\phi$
which when evaluated at $\Phi_0$ are  of higher order in slow roll.
The counterterm $\mathcal{C}_0[\Lambda,H_0]$ is independent of
$\phi$ and will be required to cancel the ultraviolet divergences in
the energy momentum tensor (see next section).

The counterterm coefficient $\mathcal{C}_2[\Lambda,H_0] $ is fixed
by requiring that it cancels the ultraviolet divergence in the
eq.(\ref{equalup}). This is achieved as follows. Introducing the
renormalized form of the inflaton potential given by eq.
(\ref{count2}) in the Lagrangian density and performing the shift in
the fields as in eq. (\ref{tad}), the Lagrangian density,
eq.(\ref{lagra}) now becomes \bea\label{lagraren}
&&\mathcal{L}[\Phi_0(t),\varphi(\vx,t)] = \frac{1}{2} \;
{\dot{\Phi}^2_0}-V_R(\Phi_0)-\delta V_R(\Phi_0)+\frac{1}{2} \;
{\dot{\varphi}^2}-\frac{(\nabla \varphi)^2}{2 \, a^2} -\frac{1}{2}
\; \left[ V^{''}_R(\Phi_0)+\mathcal{C}_2[\Lambda,H_0] \;
V^{(IV)}_R(\Phi_0)+\cdots\right] \; \varphi^2 \nonumber \\ && -
\varphi \left[ \ddot{\Phi}_0+3 \, H_0 \,
\dot{\Phi}_0+V^{'}_R(\Phi_0)+ \mathcal{C}_2[\Lambda,H_0] \;
V^{'''}_R(\Phi_0)+\cdots\right] - \frac{1}{6} \; V^{'''}_R(\Phi_0)
\; \varphi^3 - \frac{1}{24} \; V^{(IV)}_R(\Phi_0) \;
\varphi^4+\cdots \eea \noindent where the dots contain terms with
higher derivatives of the potential with respect to $\Phi_0$ which
are subleading in the slow roll expansion. In Minkowski space time
the counterterms are time independent because they must maintain the
space-time symmetries. In an spatially flat FRW cosmology only
spatial translational invariance restricts the form of the
counterterms, hence  time dependent  counterterms are allowed.

The equation of motion for $\Phi_0$ can now be obtained by
implementing the tadpole method as described above, with the leading
order result \be\label{eqofmotQ}
\ddot{\Phi}_0(t)+3\,H_0\,\dot{\Phi}_0(t)+V^{'}_R(\Phi_0)+
V^{'''}_R(\Phi_0)\left\{\mathcal{C}_2[\Lambda,H_0]+
\left(\frac{H_0}{4 \, \pi}\right)^2 \left[\Lambda_p^2+
\ln\Lambda_p^2 +\frac{1}{\Delta}+ 2 \, \gamma - 4
+\mathcal{O}(\Delta)\right] \right\}=0 \;. \ee The counterterm
$\mathcal{C}_2[\Lambda,H_0] $ is now chosen to cancel the
ultraviolet cutoff dependence in the equation of motion, namely
\be\label{countfix} \mathcal{C}_2[\Lambda,H_0]= - \left(\frac{H_0}{4
\, \pi}\right)^2 \left[\Lambda_p^2+\ln\Lambda_p^2 + 2 \, \gamma - 4
\right]\;, \ee \noindent leading to the final form of the
renormalized inflaton equation of motion to leading order in the
slow roll expansion \be\label{fineq}
\ddot{\Phi}_0(t)+3\,H_0\,\dot{\Phi}_0(t)+V^{'}_R(\Phi_0)+
\left(\frac{H_0}{4 \, \pi}\right)^2
\frac{V^{'''}_R(\Phi_0)}{\Delta}=0 \;. \ee An important aspect of
this equation  is the following: naively, the quantum correction is
of order $V^{'''}_R(\Phi_0)$, therefore  of second order in slow
roll, but the strong infrared divergence arising from the quasi
scale invariance
 of inflationary fluctuations brings about a denominator
which is of first order in slow roll. Hence the lowest order
quantum correction in the slow roll expansion,
is actually  of the same order as $ V^{'}_R(\Phi_0) $.
To highlight this observation, it proves convenient to write
eq.(\ref{fineq}) in terms of the EFT and slow roll parameters,
\be\label{fineqsr}
\ddot{\Phi}_0(t)+3\,H_0\,\dot{\Phi}_0(t)+
V^{'}_R(\Phi_0)\left[1+\left(\frac{H_0}{2\pi \, M_{Pl}}\right)^2
\frac{\xi_V}{2\,\epsilon_V\,\Delta}\right]=0 \;.
\ee
Since $\xi_V \sim \epsilon^2_V$ and $\Delta \sim \epsilon_V$ the leading quantum
corrections are of zeroth order in slow roll. This is a consequence
of the infrared enhancement resulting from the nearly scale
invariance of the power spectrum of scalar fluctuations. The quantum
correction is  suppressed by an EFT factor $H^2/M^2_{Pl} \ll 1$.

Restoring the dependence of $\Delta$ on $\Phi_0$ through the
definitions (\ref{etav}) and (\ref{delta}) we
finally find the following equation of motion for the inflaton field
in the \emph{effective field theory} up to leading order in slow
roll
\be\label{eqnslor}\ddot{\Phi}_0(t)+3\,H\,\dot{\Phi}_0(t)+
V^{'}_R(\Phi_0)+\frac{1}{24 \, (\pi \;  M_{Pl})^2} \;
\frac{V_R^3(\Phi_0)\,
V^{'''}_R(\Phi_0)}{2 \, V_R(\Phi_0)V^{''}_R(\Phi_0)-V^{'\,2}_R(\Phi_0)
}=0 \;.
\ee
We can also write the inflaton field equation in terms of
CMB observables
$$
\ddot{\Phi}_0(t)+3\,H_0\,\dot{\Phi}_0(t)+
V^{'}_R(\Phi_0)\left\{1+ \frac{\Delta_{\mathcal{R}}^2}{n_s -1 +\frac{r}4}
\left[\frac{r}2 \left(n_s -1 +\frac{3 \, r}{16} \right) -
 \frac{dn_s}{d \ln k} \right]\right\} = 0 \; .
$$

\section{Quantum corrections to the Friedmann equation:
the effective potential}\label{friedeq}

The zero temperature effective potential in Minkowski space-time is often
used to describe the scalar field dynamics during inflation
\cite{liddle,riottorev}.
The focus of this Section is to derive the effective potential for
slow-roll inflation. As we see below  the resulting effective potential
[see eq.(\ref{Veff})]  is remarkably different from the Minkowski one
[see Appendix A].

Since the fluctuations of the inflaton field are quantized, the
interpretation of the `scalar condensate' $\Phi_0$ is that of the
expectation value of the full quantum field $\phi$ in a homogeneous
coherent quantum state. Consistently with this, the Friedmann
equation must necessarily be understood in terms of the
\emph{expectation} value of the field energy momentum tensor, namely
\be\label{FRW2} H^2= \frac{1}{3 \, M^2_{Pl}}\left\langle \frac{1}{2}
\; \dot{\phi}^2+\frac{1}{2} \; \left(\frac{\nabla
\phi}{a(t)}\right)^2+V[\phi] \right\rangle  \;. \ee Separating the
homogeneous condensate from the fluctuations as in eq. (\ref{tad})
with the condition that the expectation value of the quantum
fluctuation vanishes eq.(\ref{exp}), the Friedmann equation becomes

\be\label{FRexp} H^2= \frac{1}{3 \, M^2_{Pl}}\left[ \frac{1}{2} \;
{\dot{\Phi_0}}^2 + V_R(\Phi_0)+\delta V_R(\Phi_0)\right]+ \frac{1}{3
\, M^2_{Pl}}\left\langle \frac{1}{2} \; \dot{\varphi}^2+\frac{1}{2}
\; \left(\frac{\nabla \varphi}{a(t)}\right)^2+\frac{1}{2} \;
V^{''}(\Phi_0)\; \varphi^2 +\cdots\right\rangle \ee

The dots inside the angular brackets correspond to terms with higher
derivatives of the potential which are  smaller in the slow roll
expansion. The quadratic term $\langle \varphi^2 \rangle$ has been
calculated above to leading order in slow roll and given by eq.
(\ref{QC}). Calculating the expectation value in eq.(\ref{FRexp}) in
free field theory  corresponds to obtaining the corrections to the
energy momentum tensor by integrating the fluctuations \emph{up to
one loop} (see the appendix for a similar calculation in Minkowski
space-time). We  obtain these contributions up to this order,
consistently with our study of the equations of motion up to
one-loop order.

The first two terms of the expectation value in  eq.(\ref{FRexp})
\emph{do not} feature infrared divergences for $\nu=3/2$ because of
the two extra powers of the loop momentum in the integral. These
contributions are given by \bea\label{kinterm} &&\left\langle
\frac{1}{2} \; \dot{\varphi}^2 \right\rangle = \frac{H^4_0}{16
\,\pi} \; \int^{\Lambda_p}_{0} \frac{dz}{z} \;  z^2 \;
\left|\frac{d}{dz}\left[z^{\frac{3}{2}} H^{(1)}_{\nu}(z) \right]
\right|^2 = \frac{H^4_0 \; \Lambda^4_p}{32 \,\pi^2}+
\mathcal{O}( H^4_0 \Delta)\;,\\
\label{grad} &&\left\langle\frac{1}{2}\left(\frac{\nabla
\varphi}{a(t)}\right)^2 \right \rangle = \frac{H^4_0}{16 \,\pi} \;
\int^{\Lambda_p}_{0} \frac{dz}{z} \; z^{5} \;  \left|
H^{(1)}_{\nu}(z)  \right|^2 = \frac{H^4_0 \; \Lambda^4_p}{32 \,
\pi^2}+ \frac{H^4 \; \Lambda^2_p}{16 \, \pi^2} +\mathcal{O}( H^4_0
\Delta)\;. \eea Clearly, the choice of the counterterm
$\mathcal{C}_2[\Lambda,H_0]$ given by eq. (\ref{countfix}) cancels
the ultraviolet divergences arising from the third term in the
angular brackets in eq. (\ref{FRexp}). The counterterm
$\mathcal{C}_0[\Lambda,H_0]$ in the renormalized potential
eq.(\ref{count}) is chosen to cancel the ultraviolet divergences
from the kinetic and gradient terms, namely \be\label{count0fix}
\mathcal{C}_0[\Lambda,H_0]= - \frac{\Lambda^2_p}{(4\,
\pi)^2}\left(\Lambda^2_p + H^2_0\right) \; . \ee The fully
renormalized Friedmann equation up to one loop and to lowest order
in the slow roll expansion is therefore \be\label{FRren} H^2 =
\frac{1}{3 \, M^2_{Pl}}\left[ \frac{1}{2} \; {\dot{\Phi_0}}^2 +
V_R(\Phi_0) + \left(\frac{H_0}{4 \,
\pi}\right)^2\frac{V^{''}_R(\Phi_0)}{\Delta} +\textmd{higher orders
in slow roll}\right] \equiv H^2_0 + \delta H^2 \;, \ee \noindent
where $H_0$ is the  Hubble parameter in absence of quantum
fluctuations:
$$
 H^2_0 = \frac{V_R(\Phi_0)}{3 \, M^2_{Pl}} \left[1+\frac{\epsilon_V}{3}+
\mathcal{O}(\epsilon^2_V,\epsilon_V \; \eta_V) \right] \; .
$$
Using the lowest order slow roll relation eq. (\ref{flucmass}), the
last term in eq.(\ref{FRren}) can be written as follows
\be\label{delH}
\frac{\delta H^2}{H^2_0} = \left(\frac{H_0}{4 \,
\pi\,M_{Pl}}\right)^2 \frac{\eta_V}{\Delta}\;.
\ee

This equation defines the back-reaction correction to the scale factor
arising from the quantum fluctuations of the inflaton.

Hence, while the ratio $\eta_V/\Delta$ is of order zero in slow roll,
the one loop correction to the Friedmann equation is of the order
$H^2_0/M^2_{Pl} \ll 1$ consistently with the EFT expansion.
The Friedmann equation suggests the identification of the effective potential
\bea\label{Veff}
&&V_{eff}(\Phi_0) = V_R(\Phi_0)+ \left(\frac{H_0}{4
\, \pi}\right)^2\frac{V^{''}_R(\Phi_0)}{\Delta} +\textmd{higher
orders in slow roll} = \\ \cr
&&= V_R(\Phi_0)\left[1+
\left(\frac{H_0}{4 \, \pi\,M_{Pl}}\right)^2\frac{\eta_V}{
\eta_V-\epsilon_V} + \textmd{higher orders in slow roll} \right]= \cr
\cr && =  V_R(\Phi_0)\left[1+\frac{\Delta^2_{\mathcal{T}}}{32} \;
 \frac{n_s -1 + \frac38 \; r}{n_s -1 + \frac14 \; r}
+\textmd{higher orders in slow roll}\right] \; .
\label{Vefsr}
\eea
In eq.(\ref{Vefsr}) we express the quantum corrections to the inflaton
potential in terms of observables: $ \Delta_{\mathcal{T}}, \; n_s $ and $ r $.
Using the WMAP data for $ \Delta^2_{\mathcal{R}} =
\Delta^2_{\mathcal{T}}/r = 0.218 \times 10^{-8} $, eq.(\ref{Vefsr}) becomes
$$
V_{eff}(\Phi_0) =  V_R(\Phi_0)\left[1+ 0.682 \times 10^{-10} \; r \;
 \frac{n_s -1 + \frac38 \; r}{n_s -1 + \frac14 \; r}
+\textmd{higher orders in slow roll}\right] \; .
$$
We see that the equation of motion for the inflaton eq.(\ref{fineq})
takes the natural form
$$
\ddot{\Phi}_0(t)+3\,H_0\,\dot{\Phi}_0(t)+
\frac{\partial V_{eff}}{\partial\Phi_0}(\Phi_0)
= 0 \;.
$$
where the derivative of $ V_{eff} $ with respect to $ \Phi_0 $
is taken at fixed Hubble and slow roll parameters. That is,
$ H_0 $ and $ \Delta $ must be considered in the present context
as gravitational degrees of freedom and not as matter (inflaton)
degrees of freedom.

Eqs.(\ref{delH}) and (\ref{Veff}) make manifest the nature of the
effective field theory expansion in terms of the ratio
$\left(H_0/M_{Pl}\right)^2$. The coefficients of the powers of this
ratio are obtained in the slow roll expansion. To leading order,
these coefficients are of $\mathcal{O}(\epsilon^0_V)$ because of the
infrared enhancement manifest in the poles in $\Delta$, a
consequence of the nearly scale invariant power spectrum of scalar
perturbations.

The equivalence between the (EFT) ratio $ \left(H_0/M_{Pl}\right)^2$
and the ratio $ \left(T_{sem}/T_{Pl} \right)^2$ according to
eq.(\ref{tem}), results in that the leading quantum corrections to the
effective potential eq.(\ref{Vefsr}) are $\propto T^2_{sem}$. This is
akin to the {\it finite temperature} contribution to the one loop
effective potential in  Minkowski space time.

A noteworthy result is the rather different form of the effective
potential eq.(\ref{Veff}) as compared to the result in Minkowski
space time at zero temperature.
In the appendix we show explicitly that the same
definition of the effective potential as the expectation value of
$T_{00}$ in Minkowski space-time  at zero temperature yields the familiar
one loop result, which is strikingly  different from eq.(\ref{Veff}) during
slow roll inflation.

\subsection{Quantum corrections to slow roll parameters}

We now have all of the elements in place to obtain the \emph{quantum
corrections} to the slow roll parameters. Defining the
\emph{effective} slow roll parameters as
\be\label{effSR}
\epsilon_{eff} = \frac{M^2_{Pl}}{2}\;
\left[\frac{V^{'}_{eff}(\Phi_0)}{V_{eff}(\Phi_0)}\right]^{\!
2}~~;~~\eta_{eff} =  M^2_{Pl} \;
\frac{V^{''}_{eff}(\Phi_0)}{V_{eff}(\Phi_0)} \; ,
\ee \noindent
eq.(\ref{Veff}) yields to leading order in EFT and slow roll
expansions:
\bea
&&\epsilon_{eff}  =  \epsilon_V \left[ 1+
\left(\frac{H_0}{4 \, \pi \, M_{Pl}} \right)^2 \,
\frac{4 \, \eta_V\left(\eta_V-\epsilon_V\right)-
\xi_V}{\left(\eta_V-\epsilon_V\right)^2}
\right] \label{epseff}\\
&&\eta_{eff}  =  \eta_V  \left\{1+
\left(\frac{H_0}{4 \, \pi \, M_{Pl}} \right)^2 \,
\frac1{\left(\eta_V-\epsilon_V\right)^2 }
\left[\frac{\xi_V^2}{\eta_V(\eta_V-\epsilon_V)} -
\frac{\sigma_V}{2\,\eta_V} - \frac{\xi_V}{\eta_V} \left(\eta_V+ 6 \,
\epsilon_V \right) + 4 \, \eta_V \,
\left( \eta_V + 4 \, \epsilon_V\right) - 20 \, \epsilon_V^2
\right]\right\} \; .   \nonumber
\eea
 A remarkable feature of the quantum corrections to the slow
 roll parameters is that they are of \emph{zeroth} order in slow
 roll. Again, this is a consequence of the infrared enhancement of
 the loop diagrams for a nearly scale invariant spectrum of
 fluctuations. Higher order slow roll parameters can be obtained
 similarly.

\section{Quantum corrections to superhorizon modes:  a new scaling
dimension}\label{anomdim}

In order to study the equations of motion for the fluctuations
including self-energy corrections, it is convenient to first pass to
conformal time and to implement the conformal rescaling of the field
as in eq. (\ref{rescale}). The  action is now given by
\be
S= \int d^3x \; d\eta \; \mathcal{L}_c[\chi,\Phi_0] \;,
\ee
\noindent where the Lagrangian density $\mathcal{L}_c[\chi,\Phi_0]$
is given by
\bea\label{lagconf}
&&\mathcal{L}_c[\chi,\Phi_0] =
C^4(\eta)\left[ \frac{1}{2} \; {\dot\Phi}^2_0-V_R(\Phi_0)-\delta
V_R(\Phi_0) \right] + \frac{{\chi'}^2}{2}-\frac{(\nabla
\chi)^2}{2}-\frac{1}{2} \; {\mathcal{M}^2(\eta)} \; \chi^2 - \nonumber \\
&& - C^3(\eta) \; \chi \; \left[ \ddot{\Phi}_0+3 \, H \,
\dot{\Phi}_0+V^{'}_R(\Phi_0)+\mathcal{C}_2[\Lambda,H] \;
V^{'''}_R(\Phi_0)+\cdots\right] -
\frac{1}{2} \; \delta \mathcal{M}^2(\eta) \; \chi^2
-\frac{g}{3} \; C(\eta) \; \chi^3-\frac{\lambda}{4} \; \chi^4 +\cdots
\eea
\noindent where the dots on $\Phi_0$ stand for derivatives with
respect to cosmic time, the primes on $\chi$ stand for
derivatives with respect to conformal time, and we have used the
definitions given in eqs.(\ref{g}) and (\ref{lambda}). The
effective (time dependent) mass term and counterterm are given by
\bea
&&\mathcal{M}^2(\eta) = V^{''}_R(\Phi_0) \; C^2(\eta)-
\frac{C^{''}(\eta)}{C(\eta)} =
-\frac{1}{\eta^2}(\nu^2-\frac{1}{4}) \; ,\label{renmass} \\
&&\delta \mathcal{M}^2 = \left(6 \, \lambda \;
\mathcal{C}_2[\Lambda,H_0] + g^2 \;
\mathcal{C}_3[\Lambda,H_0]\right) C^2(\eta) \label{contramasa}\;.
\eea
$ \mathcal{C}_2[\Lambda,H_0] $ is given by eq.(\ref{countfix})
and $ \mathcal{C}_3[\Lambda,H_0] $ will cancel a logarithmic
divergence proportional to $ g^2 $ in the one loop self energy.

\begin{figure}[h!]
\begin{center}
\includegraphics[height=2in, width=5in,keepaspectratio=true]{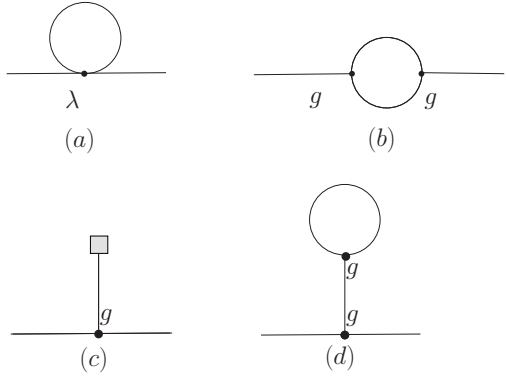}
\caption{One-loop self energy contributions.
$\lambda=\frac{1}{6} \; V^{(IV)}(\Phi_0)~,~g=\frac{1}{2} \; V^{'''}(\Phi_0)$.
The square box in diagram (c) represents $ - C^3(\eta)\chi\left[
\ddot{\Phi}_0+3 \, H \; \dot{\Phi}_0+V^{'}_R(\Phi_0)+\mathcal{C}_2[\Lambda,H]
\; V^{'''}_R(\Phi_0)\right]$.
The sum of diagrams (c) and (d) is proportional to the equation of
motion (\ref{1lupeqn}) and vanishes. Only diagrams (a) and (b)
contribute to the self-energy. } \label{selfenergy}
\end{center}
\end{figure}

The effective equation of motion for the fluctuations is obtained
in the linear response approach by introducing an external source
that induces an expectation value for the field $\chi(\vx,\eta)$,
switching off the source this expectation value will evolve in
time through the effective equation of motion of the fluctuations.
This program is implemented by following the steps explained in
our previous articles (see
ref.\cite{ultimonuestro1,ultimonuestro2} and references therein).
We first write the spatial Fourier transform of the fields
$\chi^{\pm}(\vx,\eta)$ defined on the forward $(+)$ and backward
$(-)$ branches in the generating functional, namely
\be\label{split}
\chi^{\pm}_{\vk}(\eta) = X_{\vk}(\eta) +
 \sigma^{\pm}_{\vk}(\eta)~~;~~ \langle \chi^{\pm}_{\vk}(\eta)\rangle
 =X_{\vk}(\eta) ~~;~~\langle \sigma^{\pm}_{\vk}(\eta)\rangle =0
 \;,
 \ee
 \noindent were $X_{\vk}(\eta)$ is the  spatial Fourier transform of
 the expectation value of the fluctuation field $\chi$ induced by
 the external source term. Implementing the tadpole condition
 $\langle \sigma^{\pm}\rangle=0$ up to one loop we obtain the
 effective equation of motion\cite{ultimonuestro1,ultimonuestro2}
\be \label{eqnofmotfluc}
X''_{\vk}(\eta)+\left[k^2-\frac{\nu^2-\frac{1}{4} }{\eta^2}\right]
X_{\vk}(\eta)+ \int_{\eta_0}^{\eta} \Sigma(k,\eta,\eta') \;
X_{\vk}(\eta') \; d\eta' = 0 \;.
\ee
The one-loop contributions to the self-energy kernel
$\Sigma(k,\eta,\eta')$ are displayed in fig. \ref{selfenergy}. The
sum of the diagrams $(c)$ and $(d)$ cancels by dint of the equation
of motion for the inflaton  eq. (\ref{1lupeqn}) since the loop in
diagram $(d)$ is given by $\langle [\chi(\vx,\eta)]^2 \rangle=
C^2(\eta) \; \langle [\varphi(\vx,\eta)]^2 \rangle$. Only diagrams
$(a)$ and $(b)$ give a non-vanishing contribution to the self energy
kernel, which is found to be given by
\be\label{Sigma} \Sigma(k,\eta,\eta')= \frac{1}{H^2_0 \;
\eta^2}\left[6\,\lambda\,
\left(\mathcal{C}_2[\Lambda,H_0]+\frac{1}{2} \; \langle
[\varphi(\vx,\eta)]^2\rangle\right)+ g^2 \;
\mathcal{C}_3[\Lambda,H_0]\right]\,\delta(\eta-\eta')+
\frac{2\,g^2}{H^2_0 \, \eta \, \eta'} \;
\mathcal{K}_{\nu}(k;\eta,\eta') \;. \ee The term proportional to
$\delta(\eta-\eta')$ is the sum of  the contribution from the
counterterm $\delta \mathcal{M}^2$ and the
 one loop contribution $(a)$ in fig. \ref{selfenergy}, while
the kernel $\mathcal{K}_{\nu}(k;\eta,\eta')$ is
determined by the one-loop contribution $(b)$ in fig.
\ref{selfenergy} and is given by\cite{ultimonuestro2}
\be\label{kernel}
\mathcal{K}_{\nu}(k;\eta,\eta') = 2 \int
\frac{d^3q}{(2\pi)^3} \; \mathrm{Im}\left[ S_{\nu}(q,\eta)
S^*_{\nu}(q,\eta')S_{\nu}(|\vq-\vk|,\eta)S^*_{\nu}(|\vq-\vk|,\eta')\right]\;,
\ee
\noindent where the mode functions $S_{\nu}(k,\eta)$ are given by
eq. (\ref{BDS}). The equation of motion (\ref{eqnofmotfluc}) is
solved in a  perturbative loop expansion as follows
 \be\label{pertsol}
X_{\vk}(\eta)=X_{0,\vk}(\eta)+X_{1,\vk}(\eta)+\textmd{higher loop
corrections} \;,
\ee
\noindent where $X_{0,\vk}(\eta)$ is the free field solution,
$X_{1,\vk}(\eta)$ is the one-loop correction, etc. This expansion to one loop
order leads to the following hierarchy of coupled equations
\begin{eqnarray}
&&X''_{0,\vk}(\eta)+\left[k^2-\frac{1}{\eta^2}\Big(\nu^2-
\frac{1}{4} \Big) \right]X_{0,\vk}(\eta)
 =  0  \; ,\label{X0}\\
&&X''_{1,\vk}(\eta)+\left[k^2-\frac{1}{\eta^2}\Big(\nu^2-
\frac{1}{4} \Big) \right] X_{1,\vk}(\eta) = \mathcal{R}_1(k,\eta)
\label{X1}\;,
\end{eqnarray}
\noindent where the inhomogeneity $ \mathcal{R}_1(k,\eta) $ due to
the interaction of the inflaton is given by \be\label{source}
\mathcal{R}_1(k,\eta) = - \frac{1}{H^2_0 \;
\eta^2}\left\{\frac{3\,\lambda}{2 \, \Delta}\, \left(\frac{H_0}{2 \,
\pi} \right)^2 + g^2 \; \mathcal{C}_3[\Lambda,H_0]\right\} \;
X_{0,\vk}(\eta) -\frac{ 2\, g^2 }{H^2_0 \; \eta}
\int_{\eta_0}^{\eta} \frac{d\eta'}{\eta'} \;
\mathcal{K}_{\nu}(k;\eta,\eta') \; X_{0,\vk}(\eta') \; , \ee and we
have used eqs.(\ref{QC}) and (\ref{countfix}). The solution of the
inhomogeneous eq.(\ref{X1}) is given by \be  \label{forsol}
X_{1,\vk}(\eta)= \int_{\eta_0}^{0} d\eta' \;
\mathcal{G}_\nu(k;\eta,\eta') \; \mathcal{R}_{1}(k,\eta') \; . \ee
\noindent  $\mathcal{G}_\nu(k;\eta,\eta')$ is the retarded Green's
function obeying \be\label{GF}
\left[\frac{d^2}{d\eta^2}+k^2-\frac{1}{\eta^2}\Big(\nu^2-
\frac{1}{4} \Big) \right]\mathcal{G}_\nu(k;\eta,\eta')=
\delta(\eta-\eta')~~,~~ \mathcal{G}_\nu(k;\eta,\eta')=0 \quad
\mathrm{for}\quad \eta'>\eta \; . \ee We are primarily interested in
obtaining the superhorizon behavior of the fluctuations ($ |k \,
\eta| \ll 1 $) to obtain the scaling behavior in this limit,
therefore we set  $k=0$. The kernel
$\mathcal{K}_{\nu}(0;\eta,\eta')$ was found in
reference\cite{ultimonuestro2}. To leading order in the slow roll
expansion and leading logarithmic order it is given by
\be\label{Knu}
\mathcal{K}_{\nu}(0;\eta,\eta')=\mathcal{K}_{\frac{1}{2}}(0;\eta,\eta')+
\frac{1}{6 \, \pi^2} \left[   \left(\frac{1}{2 \, \Delta}+\frac23
\right) \left(\frac{\eta'}{\eta^{2}}-\frac{\eta}{\eta^{'2}}\right) -
\frac{\eta'}{\eta^2} \; \ln\frac{\eta'}{\eta}+
\left(\frac{\eta}{\eta^{'2}}- \frac{\eta'}{\eta^2} \right) \,
\ln\left(1-\frac{\eta}{\eta'} \right) +
\frac{1}{\eta'}-\frac{1}{\eta} \right] \;,\ee \noindent where
\be\label{K12} \mathcal{K}_{\frac{1}{2}}(0;\eta,\eta')=-\frac{1}{8
\, \pi^2} \; \mathcal{P}\left( \frac{1}{\eta-\eta'}\right) =
-\frac{1}{8 \, \pi^2} \; \frac{\eta-\eta'}{\left(\eta-\eta'\right)^2
+ (\epsilon \; \eta')^2}\; . \ee \noindent $\epsilon \rightarrow 0$
furnishes a regularization of the principal part prescription in eq.
(\ref{K12}) \cite{ultimonuestro1,ultimonuestro2}. The unperturbed
solution and the retarded Green's function for $k=0$ are the
following \bea X_{0,\vec{0}}(\eta) &=&
A\;\eta^{\beta_+}+B\;\eta^{\beta_-} \quad
; \quad \beta_{\pm} \equiv \frac{1}{2}\pm \nu. \label{X00} \\
\mathcal{G}_\nu(0,\eta,\eta')&=&
\frac{1}{2\nu}\left[\eta^{\beta_+}\;{\eta'}^{\beta_-}-\eta^{\beta_-}\;
{\eta'}^{\beta_+} \right]\Theta(\eta-\eta') \;.\label{GF0} \eea
where $A$ and $B$ are arbitrary constants. Then, the last term in
eq.(\ref{source}) is \be\label{integ2} \frac{ 2\, g^2
}{H^2\;\eta}\int_{\eta_0}^{\eta} \frac{d\eta'}{\eta'} \;
\mathcal{K}_{\nu}(0;\eta,\eta') \; X_{0,0}(\eta')= A \;
\eta^{\beta_+} \; \frac{\alpha_+}{\eta^2}+B \; \eta^{\beta_-} \;
\frac{\alpha_-}{\eta^2}+F[\eta,\eta_0] \;, \ee \noindent where
$F[\eta,\eta_0]$ refers to the contribution of the lower integration
limit and does not produce secular terms in $X_{\vec{0}}(\eta)$. The
coefficients $\alpha_{\pm}$ are given by\cite{ultimonuestro2} \bea
\alpha_{\pm} & = &  \frac{g^2}{(2 \, \pi \, H_0)^2}\left[ \ln
\epsilon + \gamma + \psi(\frac12\mp\nu)\right] + \frac{g^2}{3(\pi \,
H_0)^2}\left[ \frac{1}{\frac94-\nu^2}\left(\frac{3}{2 \,\Delta} + 2
+3 \,\gamma\right) + \right. \nonumber \\ & &  \left.
+\frac{1}{\nu^2-\frac14}+\frac{1}{(\nu\pm\frac32)^2}
+\frac{\psi(\frac52 \mp \nu)}{\frac{3}{2}\mp\nu}
+\frac{\psi(-\frac12 \mp \nu)}{\frac{3}{2}\pm\nu}\right]
\;,\label{alfap} \eea Combining the above results with the first
term in eq.(\ref{source}), requiring that the counterterm
$\mathcal{C}_3[\Lambda,H_0]$ cancels the ultraviolet divergence $\ln
\epsilon $ and keeping leading terms of order $1/\Delta^2$ in
eqs.(\ref{alfap}), we find \be\label{source0} \mathcal{R}_1(0,\eta)=
A \; \frac{\eta^{\beta_+}}{\eta^2}\; (2 \,  \nu \, d^+)\,+B \;
\frac{\eta^{\beta_-}}{\eta^2} \; (2 \, \nu \,
d^-)\,+F[\eta,\eta_0]\;, \ee \noindent where the coefficients
$d^{\pm}$ of entirely quantum origin (one-loop) are given by \bea
d^+ & = & - \frac{1}{2 \, \nu}\left[\frac{3 \, \lambda}{8 \,
\pi^2\,\Delta}+ \frac{1}{6 \, \pi^2 }
\left(\frac{g}{ H_0  \, \Delta}\right)^2 \right]\label{DP} \\
d^- & = & - \frac{1}{2 \, \nu}\left[\frac{3 \, \lambda}{8 \,
\pi^2\,\Delta}+ \frac{1}{2 \, \pi^2 } \left(\frac{g}{ H_0 \,
\Delta}\right)^2\right]\;. \label{DM} \eea Carrying out the integral
in eq. (\ref{forsol}) with the retarded Green's function given by
eq. (\ref{GF0}) we find the following result for the first order
correction, \be X_{1,0}= A\;\eta^{\beta_+} \; d^+
\ln\left(\frac{\eta}{\eta_0} \right)-B\;\eta^{\beta_-} \;  d^-
\ln\left(\frac{\eta}{\eta_0}\right)+\textmd{non-secular terms}\;,
\ee \noindent where the non-secular terms  do not grow in the limit
$\eta \rightarrow 0$. Up to first order in the loop expansion and
leading order in $\Delta$, the solution for superhorizon modes is
given by \be X_0(\eta)= A\;\eta^{\beta_+}\left[1+d^+
\ln\left(\frac{\eta}{\eta_0}
\right)\right]+B\;\eta^{\beta_-}\left[1- d^-
\ln\left(\frac{\eta}{\eta_0}\right)\right]+\textmd{non-secular
terms} \;.\ee The resummation of the logarithmic secular terms is
performed by implementing the dynamical renormalization group
resummation introduced in refs.\cite{ultimonuestro1,ultimonuestro2},
leading to the following result \be\label{DRGsol}
X_0(\eta)=A_{\overline{\eta}} \;
\left(\frac{\eta}{\overline{\eta}}\right)^{\beta_++d^+} +
B_{\overline{\eta}} \;
\left(\frac{\eta}{\overline{\eta}}\right)^{\beta_--d^-}
=\left(\frac{\eta}{\overline{\eta}}\right)^{\Gamma}
\left[A_{\overline{\eta}}
\; \left(\frac{\eta}{\overline{\eta}}\right)^{\beta_++\gamma}+
B_{\overline{\eta}} \;
\left(\frac{\eta}{\overline{\eta}}\right)^{\beta_--\gamma}\right]\;,
\ee
\noindent where $\overline{\eta}$ is a renormalization scale;
the amplitudes $A_{\overline{\eta}}, \; B_{\overline{\eta}} $ are given
at this renormalization scale and obey a renormalization group
equation, so that the full solution $X_0(\eta)$ is independent of
the renormalization scale, as it must be. The exponents are given by
\bea
\gamma & = & \frac{1}{2} \left(d^++d^- \right)=
-\frac{1}{2 \, \nu}\left[ \frac{3 \, \lambda}{8 \, \pi^2\,\Delta}+
\frac{1}{3 \, \pi^2 }
\left(\frac{g}{ H_0 \,  \Delta}\right)^2 \right]\;, \label{gamma}\\
\Gamma & = & \frac{1}{2} \left(d^+-d^- \right) = \frac{1}{12 \,
\pi^2\,\nu } \left(\frac{g}{ H_0 \, \Delta}\right)^2 \;.
\label{Gamma}
\eea
The exponent $\Gamma$ coincides to leading order in slow roll with
the result  obtained in ref.\cite{ultimonuestro2}. Since $ \eta =
-e^{-H_0\,t}/H_0$, in comoving time the amplitude of superhorizon
fluctuations decays exponentially with the decay rate
 \be\label{decrat}
\Gamma_{\varphi \rightarrow  \varphi \varphi} = H_0 \, \Gamma
= \frac{H_0}{12 \, \pi^2\,\nu} \left(\frac{g}{ H_0 \, \Delta}\right)^2
\; ,
\ee
\noindent where the subscript ${\varphi \rightarrow  \varphi
\varphi}$ emphasizes that this is the rate of \emph{self decay} of
inflaton fluctuations, a novel phenomenon which is a consequence of
the inflationary expansion\cite{ultimonuestro1,ultimonuestro2}.

In ref.\cite{ultimonuestro2} no quartic coupling was considered and the
exponent $\gamma$ (for $\lambda=0$) was absorbed into a
\emph{finite} redefinition (renormalization) of the mass of the
inflaton, or alternatively of $\nu$, since the main goal of
ref.\cite{ultimonuestro2} was to obtain the decay rate $\Gamma$.
While such a redefinition of $\nu$ does not affect the decay rate
$\Gamma$, in order to understand the novel scaling dimensions $ d^{\pm} $
it must be kept separate because it originates from infrared effects and not
from the ultraviolet. The ultraviolet renormalization which is
accounted for by the counterterm $ \delta \mathcal{M}^2 $
[eq.(\ref{contramasa})], subtracts the cutoff dependent
contributions which are \emph{independent} of the wavevector $k$
(for $k \ll \Lambda$) whereas the contribution $\gamma$
[eq.(\ref{gamma})] arises from the infrared and not from the
ultraviolet, as its dependence on $\Delta$ makes manifest.
Therefore, the exponent $\gamma$ is a genuine infrared correction to
the scaling of the mode functions which cannot be absorbed by the
ultraviolet renormalization and emerges unambiguously in the scaling
regime ($k \; |\eta| \ll 1 $), i. e. physical wavelengths  much
larger than the Hubble radius. These results are  in agreement with
ref.\cite{ultimonuestro2} for the decay rate $\Gamma$. In addition
in the limit $\eta \rightarrow 0^-$ the growing mode features a
\emph{novel} scaling dimension $-d^-$ namely
\be
X_0(\eta) \buildrel{\eta \to 0}\over= B_{\overline{\eta}} \;
\left(\frac{\eta}{\overline{\eta}}\right)^{\frac{1}{2}-\nu-d^-} \;.
\ee
This correction to scaling is related to the decay rate $ \Gamma $ of
superhorizon fluctuations eq.(\ref{Gamma}). From eqs.(\ref{gcoup}),
(\ref{lam}) and (\ref{DM}), to leading order in slow roll and EFT
expansions, $ d^- $  and the comoving time decay rate
$ \Gamma_{\varphi \rightarrow \varphi\varphi}$ of superhorizon inflaton
fluctuations are given by
\bea
-d^- = && \left(
\frac{H_0}{4\pi\,M_{Pl}} \right)^{\! \! 2} \frac{\sigma_V \;
(\eta_V-\epsilon_V)+6 \, \xi^2_V}{2 \,
\epsilon_V\,(\eta_V-\epsilon_V)^2}= \Delta_{\mathcal{R}}^2 \; \frac{
\sigma_V \; (\eta_V-\epsilon_V) + 6 \, \xi^2_V}{4 \,  (\eta_V-\epsilon_V)^2}
\;, \label{diman}\\
\Gamma_{\varphi \rightarrow  \varphi\varphi} = &&  \left(
\frac{H_0}{4 \, \pi \; M_{Pl}} \right)^{\! \! 2}
\frac{H_0 \; \xi^2_V}{\epsilon_V\,(\eta_V-\epsilon_V)^2}
= \frac12 \; \Delta_{\mathcal{R}}^2 \;
\frac{H_0 \; \xi^2_V}{(\eta_V-\epsilon_V)^2} \label{gamslow}
 \eea
\noindent where the slow-roll parameters are given by eqs.
(\ref{etav})-(\ref{sig}). Whereas the exponent $\nu=\frac32-\Delta=
\frac32+\epsilon_V -\eta_V $ is determined by eq.(\ref{X0}) for the
free mode functions, the novel scaling exponent $-d^-$ is determined
by the quantum corrections  arising from the  interactions. Again
eq.(\ref{diman}) highlights an important aspect of  the effective
field theory approach. The bracket in this expression is formally of
\emph{first order} in slow roll, namely of the \emph{same} order in
slow roll as the departure from scale invariance of the \emph{free
field} mode functions. This is a consequence of the {\em infrared
enhancement} of the self-energy for $\nu \sim 3/2$ manifest as
$ \Delta^{-2}= (\eta_V-\epsilon_V)^{-2} $.  However, the
novel dimension is perturbatively small precisely because of the
effective field theory factor $H^2_0/M^2_{Pl}$.

There are two important aspects of the above results that must be
compared to our previous studies\cite{ultimonuestro1,ultimonuestro2}:
\begin{itemize}
\item{ In contrast to the study in ref.\cite{ultimonuestro1,ultimonuestro2}
we have included both the cubic and the \emph{quartic} interaction vertices.
The quartic coupling is of the same order in (EFT) as the square of the
cubic coupling but higher order in slow roll [see eqs.
(\ref{gcoup})-(\ref{lam})].
However, up to one loop, the diagram with two cubic couplings has
two propagators, while the diagram with one quartic coupling has only
one propagator. The difference in the number of propagators in these diagrams
makes the contribution from the one loop with two cubic couplings
 of the same order in (EFT) \emph{and slow roll} as the one loop with one
quartic coupling. This is an important difference with our previous study,
which a systematic treatment of \emph{both} the (EFT) and slow-roll
approximations as done here, has revealed.}

\item{In our previous study\cite{ultimonuestro1,ultimonuestro2}
the term  $\gamma = (d^+ + d^-)/2$ in the
(DRG) improved solution (\ref{DRGsol}) was absorbed into a finite mass
renormalization simply because those studies focused on the \emph{decay
rate}. Here we recognize that the contribution from $\gamma$ is \emph{not}
a mass renormalization but instead enters in the anomalous dimensions of
the growing and the decaying modes. This is an important new result,
embodied in the final expression (\ref{DRGsol}) which
displays \emph{both} the growing and decaying modes. }
\end{itemize}

\section{Inflaton coupling to other light scalars.}\label{other}

So far our analysis  only considered the self-interaction of the
inflaton. In this section we generalize the previous results to a
model which describes the  inflaton field coupled to another scalar
field $ \sigma(x) $ with both trilinear and quartic interactions. The
new action is obtained from that in  eq.(\ref{Split}) by adding the
following terms
\begin{equation}\label{2fields}
S_{\sigma}= \int d^3x \; dt \;  a^3(t) \Bigg\{ \frac{1}{2} \;
{\dot{\sigma}^2}-\frac{(\nabla \sigma)^2}{2a^2}-\frac{1}{2} \; m^2
\; \sigma^2 - g_{\sigma} \;  \phi \; \sigma^2 -
\frac{\lambda_{\sigma}}{2} \;  \phi^2 \; \sigma^2
%J(t) \; \phi + \mathrm{higher \; nonlinear \; terms}
\Bigg\}
\end{equation}
We assume initially zero expectation value for the field $\sigma$
and its time derivative. Thus, $ <\sigma> $  vanishes for all times
and inflation is still driven by one scalar field.

Upon performing the shift of the inflaton field as in eqn.
(\ref{tad}) the effective mass and trilinear coupling are given by
\bea m^2_{\sigma}(\Phi_0) & = & m^2 +
2 \, g_{\sigma} \; \Phi_0+\lambda_{\sigma} \; \Phi^2_0 \label{msigma}\\
\widetilde{g}_{\sigma}(\Phi_0) & = & g_{\sigma} +
\lambda_{\sigma} \; \Phi_0 \label{gsigma}.
\eea
In what follows we will assume that the sigma field is \emph{light}
in the sense that
\be\label{lightsig}
\frac{m^2_{\sigma}(\Phi_0)}{H^2_0} \ll 1  \; .
\ee
Now all the steps in sections III-V
generalize to this case. An extra term appears now in the equation
of motion of the inflaton (eq.\ref{1lupeqn}) at one-loop level,
\be\label{fisig} \ddot{\Phi}_0(t)+3 \, H_0 \;
\dot{\Phi}_0(t)+V'(\Phi_0)+g(\Phi_0) \; \langle
[\varphi^+(\vx,t)]^2\rangle + \widetilde{g}_{\sigma}(\Phi_0)
\;\langle [\sigma^+(\vx,t)]^2\rangle =0 \; .
\ee
We conformally rescale the $ \sigma $ field as,
$$
 \sigma(\vx,t) =\frac{\rho(\vx,\eta)}{C(\eta)} \; .
$$
The spatial Fourier transform of the free field Heisenberg
operators $\rho(\vx,\eta)$ obey the equation
\be\label{ecro}
 \rho^{''}_{\vk}(\eta)+ \left[k^2 + m^2_{\sigma}(\Phi_0) \;  C^2(\eta)-
\frac{C^{''}(\eta)}{C(\eta)} \right]\rho_{\vk}(\eta)=0 \,.
\ee
Using the slow roll expressions eq.(\ref{quasiDS}), it becomes
\be\label{ecro2} \rho^{''}_{\vk}(\eta)+ \left[k^2
-\frac{\bnu^2-\frac{1}{4}}{\eta^2} \right]\rho_{\vk}(\eta)=0 \ee
\noindent where the index $\bnu$ is given by \be\label{nub} \bnu =
\frac{3}{2} + \epsilon_V - \frac{m^2_{\sigma}(\Phi_0)}{3 \, H^2_0}
+\mathcal{O}\left[\epsilon^2_V,\eta^2_V,\epsilon_V\eta_V,
\frac{m^4_{\sigma}(\Phi_0)}{
\, H^4}\right] \,.
\ee
The parameter $ {\bar\Delta} $ which controls the infrared behavior
of the $\rho$ fluctuations is given by
\be\label{delba}
{\bar\Delta}\equiv \frac32 - \bnu =
\frac{m^2_{\sigma}(\Phi_0)}{3 \, H^2_0}-\epsilon_V
+\mathcal{O}\left[\epsilon^2_V,\eta^2_V,\epsilon_V\eta_V,
\frac{m^4_{\sigma}(\Phi_0)}{
\, H^4_0}\right] \; .
\ee
Notice that the CMB anisotropy observations indicate that the slow
roll parameters are  $  \lesssim 10^{-2} $ consequently $\Delta$  is
also  $ \lesssim 10^{-2} $.  The validity of the slow roll
approximation and the condition that the scalar field $\sigma$ be
light (\ref{lightsig}) guarantee that $ {\bar\Delta} \ll 1 $.

The renormalized effective equation of motion (\ref{1lupeqn})
becomes \be\label{fineqsig}
\ddot{\Phi}_0(t)+3\,H_0\,\dot{\Phi}_0(t)+V^{'}_R(\Phi_0)+
\left(\frac{H_0}{4 \, \pi}\right)^2
\left[\frac{V^{'''}_R(\Phi_0)}{\Delta}+ \frac{2 \,
\widetilde{g}_{\sigma}(\Phi_0)}{{\bar\Delta}}\right]=0 \;. \ee There
are  now extra diagrams that contribute to the inflaton self-energy
$ \Sigma(k,\eta,\eta') $ [eq.(\ref{Sigma})] with the field $ \sigma
$ in the internal lines.  These yield further corrections of order $
\widetilde{g}_{\sigma}^2 $ and $\lambda_{\sigma}$ to the inflaton
mode functions $ X_{\vk}(\eta) $ through eq.(\ref{eqnofmotfluc}).
The calculation of these new contributions to $ X_{\vk}(\eta) $ is
straightforward since the diagram 1 (b) with the field $ \sigma $ in
the internal loop yields the kernel
$\mathcal{K}_{\bnu}(k;\eta,\eta') $ which follows from $
\mathcal{K}_{\nu}(k;\eta,\eta') $ [eqs.(\ref{kernel} ) and
(\ref{Knu})] by simply replacing $ \nu $ by $ \bnu $ and accounting
properly for the different combinatorial factors in the Feynman
diagrams. We obtain the following contributions to the coefficients
$d^{\pm}$ from the self-energy correction that features one loop of
the $\sigma$ field \bea d^+_{\sigma} & = & - \frac{1}{2 \, \nu \,
{\bar\Delta}}\left\{\frac{ \lambda_{\sigma}}{8 \, \pi^2}+ \frac{1}{6
\, \pi^2 \, \Delta} \left[\frac{\widetilde{g}_{\sigma}(\Phi_0)}{
H_0}\right]^2 \right\}\cr \cr d^-_{\sigma} & = & - \frac{1}{2 \, \nu
\, {\bar\Delta}} \left\{\frac{ \lambda_{\sigma}}{8 \, \pi^2}+
\frac{1}{2 \, \pi^2 \, \Delta}
\left[\frac{\widetilde{g}_{\sigma}(\Phi_0)}{H_0}\right]^2\right\}\;.
\label{DMS} \eea to leading order for small $ {\bar\Delta}  \ll 1 $
and $ \Delta \, \ll 1 $.

The contribution from the light $\sigma$ field to the inflaton
self-energy leads to a further correction to the scaling dimension,
given by $-d^-_{\sigma}$, and to the decay given by
\be\label{Gamsig} \Gamma^{\sigma}= \frac{1}{2}
\left(d^+_{\sigma}-d^-_{\sigma}\right)=
\frac{1}{12 \, \pi^2 \; \nu\Delta \; {\bar\Delta}}
\left[\frac{\widetilde{g}_{\sigma}(\Phi_0)}{
H_0}\right]^2
\ee
In comoving time the partial rate of superhorizon inflaton
fluctuations to decay in a pair of $\sigma$ scalars is given by
\be\label{sigwidth}
\Gamma_{\varphi\rightarrow \sigma\sigma} =
\frac{H_0}{12 \, \pi^2 \; \nu \; \Delta \; {\bar\Delta}}
\left[\frac{\widetilde{g}_{\sigma}(\Phi_0)}{H_0}\right]^2
\ee
$ \Gamma_{\varphi\rightarrow \sigma\sigma} $ has a similar structure
to the inflaton self-coupling decay $ \Gamma_{\varphi \rightarrow
\varphi \varphi} $ [eq.(\ref{decrat})].
The \emph{total} rate of superhorizon fluctuations is given by
\be\label{Gamtot}
\Gamma_{tot} = \Gamma_{\varphi \rightarrow
\varphi\varphi}+\Gamma_{\varphi\rightarrow \sigma\sigma} =
\frac{1}{12 \, \pi^2 \; \nu \; \Delta  \; H_0}
\left[\frac{\widetilde{g}_{\sigma}^2(\Phi_0)}{{\bar\Delta}}
+ \frac{g^2(\Phi_0)}{\Delta}\right] \; ,
\ee
\noindent where $\Gamma_{\varphi \rightarrow  \varphi\varphi}$ is
given by eq. (\ref{gamslow}).

Depending on the values of the effective couplings and mass
$\widetilde{g}_{\sigma}(\Phi_0)$, $\lambda_{\sigma}$,
$m^2_{\sigma}(\Phi_0)$ the value of $d^{\pm}_{\sigma}$ can exceed
that of  $ d^{\pm} $. The value of the couplings of the inflaton to
another scalar field must necessarily be small so that there is no
large isocurvature perturbation contribution.

\section{Conclusions and further questions}\label{conclu}

Motivated by the current and forthcoming precision CMB data,  we
study the quantum corrections to the inflationary dynamics arising
from inflaton \emph{self-interactions}. Our approach distinctly
treats inflationary dynamics in terms of scalar fields as a
renormalized \emph{effective field theory} (EFT) valid when the
scale of inflation is much smaller than the cutoff scale (presumably
the Planck scale). We focus on single field slow-roll inflation as a
viable model that provides a generic setting for the robust
predictions of inflation: almost scale invariant spectrum of
gaussian adiabatic perturbations, small ratio of tensor to scalar
amplitudes, etc, which is compatible with the WMAP data. The
calculations of inflationary parameters within the slow roll
expansion are typically based on purely gaussian quantum
fluctuations. However, the WMAP data yields a hint of a cubic
interaction in the form of the small but non-vanishing `jerk'
parameter $ \xi_V$. Furthermore, self-interactions of fluctuations
will necessarily be present if the inflaton potential is non-linear
as would be the most natural effective description. In this article
we focused in obtaining the lowest order quantum corrections to the
following relevant inflationary ingredients:

\begin{itemize}
\item{ The equation of motion for the homogeneous expectation
value of the inflaton field.}
\item{The Friedmann equation. The
corrections to the Friedmann equation correspond to the quantum
contributions in the \emph{effective potential} since these are
interpreted as the expectation value of $T_{00}$ in an homogeneous
coherent state. The quantum corrections to the effective potential
yield \emph{quantum corrections} to the slow
roll parameters. These are computed to leading order in EFT and
slow roll expansions.  }
\item{The equations of motion for the
quantum fluctuations of the inflaton around its classical value.}
\end{itemize}

The effective field theory approach requires that the ratio of the
scale of inflation to the cutoff scale is small, namely $H/M_{Pl}\ll
1$. Slow roll inflation relies on an \emph{adiabatic} evolution of
the scalar field and implies a hierarchy of small dimensionless
parameters which are related to derivatives of the scalar potential
with respect to the field\cite{barrow,liddle}. We combined both
approaches to obtain the quantum corrections in the effective field
theory to leading order in the (EFT) and slow roll expansions.

Both the one-loop correction to the equation of motion of the
inflaton and to the Friedmann equation are determined by the power
spectrum of scalar fluctuations. A nearly scale invariant spectrum
entails a strong infrared behavior of the loop integral manifest as
\emph{poles} in the small parameter  $\Delta= \eta_V-\epsilon_V+
\mathcal{O}(\epsilon^2_V,\eta^2_V,\epsilon_V\eta_V)$, which measures
the departure from scale invariance.

This slow roll parameter provides a natural \emph{infrared} regularization
of the loop integrals which allows a controllable and systematic (EFT)
and slow roll expansions.

We find that the one-loop effective potential to the leading order
in slow-roll and $\Delta$ expansions is given by
\bea &&
V_{eff}(\Phi_0)= V_R(\Phi_0)\left[1+ \left(\frac{H_0}{4 \, \pi \;
M_{Pl}}\right)^2\frac{\eta_V}{ \eta_V-\epsilon_V} + \textmd{higher
orders in slow roll} \right] = \cr \cr &&=
V_R(\Phi_0)\left[1+\frac{\Delta^2_{\mathcal{T}}}{32} \frac{n_s -1 +
\frac38 \; r}{n_s -1 + \frac14 \; r} +\textmd{higher orders in slow
roll}\right] \; ,
\eea \noindent
where $H_0$ and the slow roll
parameters are explicit but slowly varying functions of $\Phi_0$
given by eqs.(\ref{etav})-(\ref{sig}). This result is strikingly
different from that in Minkowski space-time (see appendix).

>From this effective potential we obtain the effective slow roll
parameters that include \emph{quantum corrections}. The lowest
effective slow roll parameters are given by eqs.(\ref{epseff}).
A remarkable aspect of the \emph{quantum
corrections} to the effective potential and slow roll parameters
(namely the terms of order $H^2_0/M^2_{Pl}$) is that these are of
\emph{zeroth} order in slow roll. This is a consequence of the
infrared enhancement from the nearly scale invariant scalar quantum
fluctuations.

The equations of motion for the quantum fluctuations of the inflaton
around the  $\Phi_0$ are obtained including the one-loop self-energy
corrections. The self-energy  features a strong infrared behavior
which is  regularized by $\Delta$. The perturbative solution of the
equations of motion features secular terms that diverge in the long
time limit (vanishing conformal time). The dynamical renormalization
group program\cite{ultimonuestro1,ultimonuestro2} is implemented to
provide a resummation of this perturbative series. The improved
solution for superhorizon modes feature two novel aspects:  a
correction to the scaling exponent and a decay rate. The latter, as
anticipated in refs.\cite{ultimonuestro1,ultimonuestro2} describes
the \emph{self-decay} of inflaton fluctuations. To leading order in
the effective field theory and slow-roll expansions we find that the
novel scaling dimension and decay rate are given by \bea \label{dG}
-d^- = && \left( \frac{H_0}{4\pi\,M_{Pl}} \right)^{\! \! 2}
\frac{\sigma_V \; (\eta_V-\epsilon_V)+6 \, \xi^2_V}{2 \,
\epsilon_V\,(\eta_V-\epsilon_V)^2}\; , \cr \Gamma_{\varphi
\rightarrow \varphi \varphi} = && \left( \frac{H_0}{4\pi\,M_{Pl}}
\right)^{\! \! 2} \frac{H_0 \;
\xi^2_V}{\epsilon_V\,(\eta_V-\epsilon_V)^2}. \eea

Both the novel scaling dimension and the decay rate can be expressed
in terms of CMB observables inserting eqs.(\ref{gorda}) for $
\eta_V, \; \epsilon_V , \;\xi_V $ and $ \sigma_V $ into
eqs.(\ref{dG}).

While the quantum corrections  are  small, consistently with the
effective field theory expansion, they may \emph{compete} with
higher order slow roll corrections in the gaussian approximation.
Therefore,  in order to gain understanding of the inflationary
parameters from high precision data, any high order estimate in the
slow roll approximation must be accompanied by an assessment of the
quantum corrections arising from the interactions as studied here.

 We have generalized these results by studying a model in which the inflaton
is coupled to a light scalar
 field $\sigma$. We have  obtained the contribution from a loop of
$\sigma$ particles to the effective
 potential, scaling exponents and the partial rate for the decay of
superhorizon fluctuations of the inflaton
 $\rightarrow \sigma \sigma$.

In this article we have focused on studying the effects of the
interaction on the expectation value and the fluctuations of the
inflaton field. In order to understand the possible effects of the
interaction on curvature perturbations and or gravitational waves,
the next stage  of the program requires to quantize the
perturbations and to compute loop corrections. This will be the
focus of forthcoming studies.

\begin{acknowledgments} D.B.\ thanks the US NSF for support under
grant PHY-0242134,  and the Observatoire de Paris and LERMA for
hospitality during this work. This work is supported in part by the
Conseil Scientifique de l'Observatoire de Paris through an `SInAction
Initiative'.
\end{acknowledgments}

\appendix

\section{One loop effective potential and equations of motion in
Minkowski space-time: a comparison}

In this appendix we establish contact with familiar effective
potential both at the level of  the equation of motion for the
expectation value of the scalar field, as well as the expectation
value of $T_{00}$.

In Minkowski space time the spatial Fourier transform of the field
operator is given by
\be \varphi_{\vk}(t) = \frac{1}{\sqrt{2\omega_k}}\left[ a_{\vk} \;
e^{-i\omega_k\,t} + a^{\dagger}_{-\vk} \;
e^{i\omega_k\,t}\right]\,, \ee
\noindent where the vacuum state is annihilated by $a_{\vk}$ and the
frequency is given by
\be
\omega_k = \sqrt{k^2+V^{''}(\Phi_0)} \,.
\ee
The one-loop contribution to the equation of motion (\ref{1lupeqn})
is given by
\be\label{1lupmink}
\frac{V^{'''}(\Phi_0)}{2} \; \langle
[\varphi(\vx,t)]^2\rangle = \frac{V^{'''}(\Phi_0)}{8  \, \pi^2}
\int^{\Lambda}_0 \frac{k^2}{\omega_k} \; dk = \frac{d}{d\Phi_0}
\left[ \frac{1}{4\pi^2} \int^{\Lambda}_0 k^2 \; \omega_k \; dk
\right] \,.
\ee
The expectation value of $T_{00}=\mathcal{H}$ (Hamiltonian density)
in Minkowski space time is given up to one loop by the following
expression
\be \langle T_{00} \rangle = \left\langle
\frac{1}{2} \; \dot{\phi}^2+\frac{1}{2} \; \left(\nabla \phi
\right)^2+V(\phi) \right\rangle =  \frac{1}{2} \; {\dot{\Phi_0}}^2 +
V(\Phi_0)+ \left\langle
\frac{1}{2} \; \dot{\varphi}^2+\frac{1}{2} \; \left({\nabla
\varphi}\right)^2+\frac{1}{2} \; V^{''}(\Phi_0) \; \varphi^2
+\cdots\right\rangle \;. \ee
The expectation value of the fluctuation contribution  is given by
\be \left\langle
\frac{1}{2} \; \dot{\varphi}^2+\frac{1}{2} \; \left({\nabla
\varphi}\right)^2+\frac{1}{2} \; V^{''}(\Phi_0) \; \varphi^2
+\cdots\right\rangle = \frac{1}{4\pi^2} \int^{\Lambda}_0
k^2  \; \omega_k \; dk  =
\frac{\Lambda^4}{16\pi^2}+\frac{V^{''}(\Phi_0) \; \Lambda^2}{16 \,\pi^2}-
\frac{[V^{''}(\Phi_0)]^2}{64 \, \pi^2}
\ln\frac{4 \,\Lambda^2}{V^{''}(\Phi_0)} \;.
\ee
Renormalization proceeds as usual by writing the bare Lagrangian in
terms of the renormalized potential and counterterms. Choosing the
 counterterms to cancel the quartic,
quadratic and logarithmic ultraviolet divergences, we obtain the
familiar renormalized one loop effective potential
\be \label{potefM}
V_{eff}(\Phi_0) = V_R(\Phi_0)+\frac{[V^{''}_R(\Phi_0)]^2}{64 \, \pi^2}
\ln\frac{V^{''}_R(\Phi_0)}{M^2} \; ,
\ee
\noindent where $M$ is a renormalization scale.  Furthermore from
eq. (\ref{1lupmink}) it is clear that the equation of motion for
the expectation value is given by
\be\label{eqnmink}
\ddot{\Phi}_0+ V^{'}_{eff}(\Phi_0)= 0 \;.
\ee

\end{document}